\documentclass[]{aa}  

\newcommand{\chisq} {$\chi^2_\mathrm{red}$\xspace}
\newcommand{\Msun}{\>{$\rm M_{\odot}$}\xspace}
\newcommand{\Rsun}{\mbox{$\rm R_{\odot}$}}

\newcommand{\num} {$\nu_{\rm max}$\xspace}

\newcommand{\dnu} {$\Delta\nu$\xspace}
\newcommand{\TESS} {TESS\xspace}

\newcommand{\Gaia} {\textit{Gaia}\xspace}

\newcommand{\GaiaDR} {\textit{Gaia}\,DR3\xspace}

\newcommand{\Kepler} {\textit{Kepler}\xspace}

\newcommand{\KIC}[1]{{KIC\,#1\xspace}}

\newcommand{\muHz}{\,\textmu Hz\xspace}
\newcommand{\dex}{\,dex\xspace}

\def\dnu{$\Delta\nu$\xspace}
\def\dn1{$\delta\nu_{01}$\xspace}
\def\dn2{$\delta\nu_{02}$\xspace}

\def\KIC{KIC\,9163796\xspace}

\usepackage{xcolor}

\usepackage{graphicx}
\usepackage{txfonts}
\usepackage{listings}
\usepackage{xspace}
\usepackage{multirow,bigdelim}
\usepackage[colorlinks=true,     linkcolor=blue,  filecolor=blue, urlcolor=blue,citecolor=blue]{hyperref} 
\begin{document}
\defcitealias{Beck2018}{BKP18}

\title{Improving the stellar age determination through \\ joint modeling of binarity and asteroseismology}

\subtitle{Grid modeling of the seismic red-giant binary KIC\,9163796}

\titlerunning{Improving the stellar age determination through asteroseismic and binary grid modeling}
\authorrunning{D.\,H. Grossmann}

\author{D.\,H. Grossmann\inst{\ref{inst:IAC},\ref{inst:ULL}} 
\and P.\,G. Beck\inst{\ref{inst:IAC},\ref{inst:ULL}}
\and S.\,Mathur\inst{\ref{inst:IAC},\ref{inst:ULL}} 
\and C.\,Johnston \inst{\ref{inst:MPIA},\ref{inst:Rad},\ref{inst:KU}} 
\and D.\,Godoy-Rivera \inst{\ref{inst:IAC},\ref{inst:ULL}}
\and \\
J.\,C.\,Zinn  \inst{\ref{inst:Joel}}, S.\,Cassisi \inst{\ref{inst:Santi},\ref{inst:Santi2}} 
\and 
B.\,Liagre \inst{\ref{inst:CEA}} 
\and T.\,Masseron\inst{\ref{inst:IAC},\ref{inst:ULL}}\and
R.\,A.\,Garc\'ia \inst{\ref{inst:CEA}} 
\and 
A.\,Hanslmeier \inst{\ref{inst:Graz}}
\and \\
N.\,Muntean \inst{\ref{inst:Graz}}
\and L.\,S.\,Schimak \inst{\ref{inst:Sydney}}
\and L.\,Steinwender \inst{\ref{inst:TUGraz}} 
\and D.\,Stello \inst{\ref{inst:Sydney},\ref{inst:UNSW}}
} 

\institute{Instituto de Astrof\'{\i}sica de Canarias, E-38200 La Laguna, Tenerife, Spain; \label{inst:IAC}
\email{desmond.grossmann@iac.es}
\and Departamento de Astrof\'{\i}sica, Universidad de La Laguna, E-38206 La Laguna, Tenerife, Spain \label{inst:ULL}
\and Max-Planck-Institut für Astrophysik, Karl-Schwarzschild-Straße 1, 85741 Garching bei München, Germany \label{inst:MPIA}
\and Department of Astrophysics, IMAPP, Radboud University Nijmegen, PO Box 9010, 6500 GL Nijmegen, The Netherlands \label{inst:Rad}
\and Institute of Astronomy, KU Leuven, Celestijnenlaan 200D, 3001 Leuven, Belgium \label{inst:KU}
\and Department of Physics and Astronomy, California State University, Long Beach, Long Beach, CA 90840, USA \label{inst:Joel}
\and INAF - Osservatorio Astronomico d'Abruzzo, Via M. Maggini sn., Teramo, Abruzzo, Italy \label{inst:Santi}
\and INFN - Sezione di Pisa, Largo Pontecorvo 3, 56127 Pisa, Italy \label{inst:Santi2}
\and Universit\'e Paris-Saclay, Universit\'e Paris Cit\'e, CEA, CNRS, AIM, 91191, Gif-sur-Yvette, France \label{inst:CEA}
\and Institut für Physik, Karl-Franzens Universität Graz, Universitätsplatz 5/II, NAWI Graz, 8010 Graz, Austria \label{inst:Graz}
\and Sydney Institute for Astronomy (SIfA), School of Physics, University of Sydney, NSW 2006, Australia \label{inst:Sydney}
\and Graz University of Technology, Rechbauerstraße 12, Graz, Austria \label{inst:TUGraz}
\and School of Physics, University of New South Wales, NSW 2052, Australia  \label{inst:UNSW}
}

\date{Submitted on October 15, 2024. Accepted on January 14, 2025.}
 
  \abstract
   {Typical uncertainties of ages determined for single star giants from isochrone fitting using single-epoch spectroscopy and photometry without any additional constraints are 30-50\,\%. Binary systems, particularly double-lined spectroscopic binaries, provide an opportunity to study the intricacies of internal stellar physics and better determine stellar parameters, particularly the stellar age.}
   {By using the constraints from binarity and asteroseismology, we aim to obtain precise age and stellar parameters for the red giant-subgiant binary system \KIC, a system with a mass ratio of 1.015 but distinctly different positions in the Hertzsprung–Russell diagram.
   }
   {We compute a multidimensional model grid of individual stellar models. From different combinations of figures of merit, we use the constraints drawn from binarity, spectroscopy, and asteroseismology to determine the stellar mass, chemical composition, and age of \KIC. 
   }
   {Our combined-modeling approach leads to an age estimation of the binary system \KIC
   of 2.44\,$^{+0.25}_{-0.20}$\,Gyr, which corresponds to a relative error in the age of 9\,\%. Furthermore, we found both components exhibiting equal initial helium abundance of 0.27 to 0.30, significantly higher than the primordial helium abundance, and an initial heavy metal abundance below the spectroscopic value. The masses of our models are in agreement with masses derived from the asteroseismic scaling relations. }
   {By exploiting the unique, distinct positions of \KIC, we successfully demonstrated that combining asteroseismic and binary constraints leads to a significant improvement of precision in age estimation, that have a relative error below 10\% for a giant star. }

\keywords{Asteroseismology 
$-$ (Stars:) binaries: spectroscopic
$-$ Stars: late-type $-$ Stars: oscillations (including pulsations), Stars:~individual:~KIC\,9163796.}

\maketitle
%

\nolinenumbers 

\section{Introduction
\label{sec:introduction}}

Knowing the precise ages of stars is crucial for many fields of astrophysics \citep{Agereview}, in particular galactic archaeology \citep{Freeman2002} and stellar population studies \citep{Wang2024}. A commonly used technique for stellar age determination is isochrone fitting. However, it only applies with reasonable precision to stars close to the subgiant turnoff but can show large uncertainties for main-sequence field stars \citep{Valle2013,Lebreton2014helium,GodoyRiviera2021}. For red giants, this method delivers relative age uncertainties of 30-50\,\% \citep{Cassagrande2016}. In the last decade, with the advent of space missions such as NASA's \Kepler \citep{Borucki2010} and Transiting Exoplanet Survey Satellite  \citep[TESS, ][]{Ricker2014}, asteroseismology has gained significance as a tool for parameter determination including age for different types of stars, in particular, red giant branch (RGB) stars \citep[e.g.][]{LiTanda2022,WarfieldApocasc2024}. Generally, two approaches for asteroseismic age determination exist: modeling using the frequency of maximum oscillation power \num and the large frequency separation \dnu \citep[i.e., the global seismic parameters; e.g.,][]{Valle2015a,Cassagrande2016,APOKASC2}, or boutique modeling of oscillation modes \citep[e.g.][]{Lebreton2014helium,Campante2023,LiTanda2024}. While the latter can deliver precise results for certain targets, calculating comprehensive frequency grids for post-main-sequence stars, in particular RGB and red-clump stars, is very resource intensive, and the modes from models are susceptible to poorly-modeled surface effects, particularly in active stars \citep[e.g.][]{Garcia2014b,Salabert2016}. Ages from modeling using the global seismic parameters can be especially powerful if combined with additional constraints, such as in the case of globular clusters \citep[e.g.][]{Moser2023,Howell2024}, open clusters \citep{Brogaard2023} or binary systems \citep{Johnston20192,Murphy2021}. 

Depending on the spectral type and age, estimations are that 50\,\% to 100\,\% of stars are in binary systems \citep[e.g.,][and references therein]{Raghavan2010,MoeStefano2017, Badenes2018, Offner2023}. Unless the binary system was created from a very rare capturing event, the components of binary systems were born in the same cloud at the same time \citep[e.g.,][]{Prsa2018,MoeStefano2017}. Therefore, they exhibit identical initial conditions, such as initial chemical abundances, ages, and distance to the observer. These binary constraints enable us to use such coeval systems as test beds for stellar physics and evolution \citep[e.g.][]{Cassisi2011,Burgo2018,Valle2018,Valle2023a,Johnston2019,Murphy2021}. Especially powerful is the combination of a binary system with at least one of its components exhibiting solar-like oscillations. Until recently, only around 100 of those unique systems were known \citep{Beck2022}. It was only the third data release \citep[DR3; ][]{GaiaDR3} of the ESA \Gaia mission \citep{GaiaMain}, which included a dedicated non-single-star catalog for the first time, that made it possible to increase the number of known oscillating binary systems up to 1000 \citep{BeckGrossmann2023}. However, only very few red-giant systems with both components exhibiting a power excess have been studied \citep{Rawls2016,Beck2018}.
\begin{figure}[t!]
    \centering
    \includegraphics[width=\columnwidth
    ]{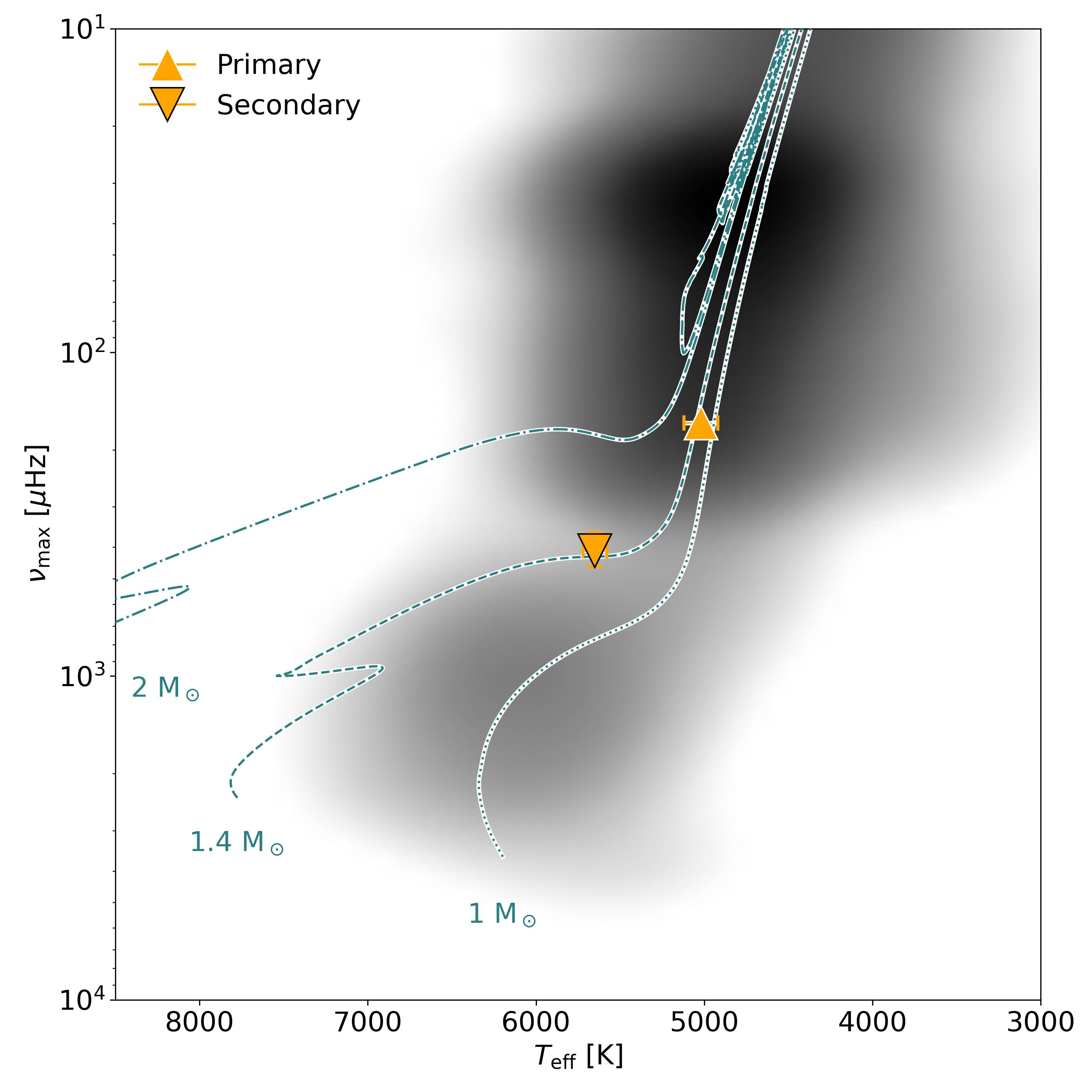}
    \caption{HRD depicting the positions of the primary (upward triangle) and secondary (downward triangle) component of \KIC according to the asteroseismic analysis from \cite{Beck2018}. As references, the tracks of a 1 \Msun, 1.4 \Msun and 2 \Msun evolutionary track with the primary's metallicity from MESA are provided as dashed lines. The background contour plot shows the density distribution of all targets from the input catalogs used in \cite{BeckGrossmann2023, Mathur2022} with an asteroseismic detection.}\label{fig:HRD} 
    
\end{figure}
One of these previously studied systems is \KIC, a double-lined spectroscopic binary system (SB2) composed of a red-giant branch (RGB) star and a sub-giant star, that has been described observationally in detail by \citet[hereafter BKP18]{Beck2018}. Detections of oscillations in both components from \Kepler light curves, combined with its SB2 nature, enabled \citetalias{Beck2018} to give masses and radii for both components.  Given its binary nature, the differences in the observational parameters for the two stars, described in detail in Sect.\,\ref{sec:observations}, are solely caused by the difference in mass between the components of 1.5\,$\pm$\,0.5\% \citepalias{Beck2018}. This mass difference also causes the two components to be in distinct and favorable positions on the Hertzsprung-Russel Diagram (HRD), as shown in Fig.\,\ref{fig:HRD}. The primary is moving up the RGB and is strongly affected by changes in the mixing length parameter in modeling while exhibiting small fractional changes in effective temperature.  On the contrary, the subgiant exhibits large fractional changes in effective temperature per unit time as it moves horizontally in very short timescales in the HRD. Due to its observations with spectroscopy, asteroseismology, and binarity, and the component's distinct positions in the HRD (Fig.\,\ref{fig:HRD}), \KIC enables us to break degeneracies, such as the mass-initial helium degeneration, and obtain a precise age estimation for the system.

Here, we present an asteroseismic grid modeling and age determination of the binary system \KIC by combining models from stellar evolution and oscillation codes with observational constraints from spectroscopy, asteroseismology, and binarity.
The paper is structured as follows.
In Sect.\,\ref{sec:observations}, we provide an overview of observational constraints of \KIC in the literature, which are relevant for the modeling. 
Section\,\ref{sec:grid} explains how the model grid was obtained, while Sect.\,\ref{sec:bestfit} describes the figure of merit and error estimation used for modeling. In Sect.\,\ref{sec:cases} we describe the search for the best-fitting model combinations for the different cases of the figure of merit and different constraints from binarity. Sect.\,\ref{sec:discussion} discusses the results from this grid modeling approach. We conclude in Sect.\,\ref{sec:conclusion}. 
\begin{table*}[t!]
\caption{Summary of observational parameters of \KIC.}
\tabcolsep=10pt 

\centering
\begin{tabular}{lccc}
\hline
\hline
Parameter                                    &                                          & Primary                            & Secondary                      \\ \hline
Mass ratio $q=M_1/M_2$                       &                                          & \multicolumn{2}{c}{$1.015\pm0.005$}                                 \\
Metallicity {[}Fe/H{]}                       & [dex]                                      & $-0.37$\,$\pm$\,$0.1$                      & $-0.38$\,$\pm$\,$0.1$                  \\
Effective temperature $T_{\mathrm{eff}}$     & {[}K{]}                                  & 5020\,$\pm$\,100                       & 5650\,$\pm$\,70                    \\
Surface gravity $\log_g$     & [dex]                & $3.14$\,$\pm$\,$0.2$                           & $3.48$\,$\pm$\,$0.3$                                             \\
\hline
Frequency of maximum oscillation $\nu_{max}$ & \multicolumn{1}{l}{[\muHz]} & \multicolumn{1}{l}{165.3\,$\pm$\,1.3} & 340\,$\pm$\,20\\
Large frequency separation \dnu                  & \multicolumn{1}{l}{[\muHz]} & \multicolumn{1}{l}{12.85\,$\pm$\,0.03} & $-$                            \\
Mass M                                       & {[}\Msun{]}                          & $1.39$\,$\pm$\,$0.06$                      & $-$                            \\ \hline 
Magnitude difference  $\Delta$\,m                       & \multicolumn{1}{l}{[mag]}                  & \multicolumn{2}{c}{0.58\,$\pm$\,0.08}                                   \\
Gaia mean magnitude $G_{mean}$               & [mag]                                      & \multicolumn{2}{c}{9.6}                                             \\ \hline
\end{tabular}
\label{tab:observables}
\tablefoot{All observational data presented, grouped in spectroscopic, seismic, and photometry, have been published in \citetalias{Beck2018}, except the \GaiaDR mean magnitude \citep{GaiaDR3}.}
\end{table*}

\section{Observational constraints of the system \label{sec:observations}}

\KIC \footnote{\KIC = TIC 271664383, \GaiaDR 2079986695058294272, and 2MASS J19412099+4530173 } was shown by \cite{Beck2014} to be an oscillating red-giant sub-giant binary located in an eccentric orbit ($e$\,=\,0.69\,$\pm$\,0.01) with an orbital period of P$_\mathrm{orb}$\,=\,121.30\,$\pm$\,0.01\,d. 
The system was initially detected in \Kepler photometry due to a periodic flux variation induced by tides during periastron passage in a highly eccentric orbit \citep{Zahn1975, Remus2012}. Due to the resemblance of echo-cardiograms in their lightcurve \cite{Thompson2012} coined the 
phrase 'heartbeat stars' for this class of eccentric binaries \citep{Kumar:1995, Welsh:2011}. 

The original analysis by \cite{Beck2014} revealed solar-like oscillations with relatively low amplitudes compared to similar red giants within the bulk asteroseismic sample \citep[e.g.][]{Kallinger2014}. In a detailed study of the system, \citetalias{Beck2018} presented a revised asteroseismic analysis based on the $\sim$1400\,days long of the \Kepler light curve (\Kepler Quarters 0-14). The seismic analysis was done on the \texttt{KEPSEISMIC}\footnote{\href{https://archive.stsci.edu/prepds/kepseismic/}{https://archive.stsci.edu/prepds/kepseismic/}} data  \citep[for details of the database and calibration see][]{Garcia2011,Garcia2014a, Pires2015}. By re-evaluating the seismic parameters and spectroscopic parameters from \citealt{Beck2014} (Revised values: \num = {165.3\,$\pm$\,1.3\,\muHz}; \dnu\,=\,$12.83$\,$\pm$0.03\,\muHz; $T_{\rm eff}$=5020 K), \citetalias{Beck2018} determined a revised estimate for the seismic mass and radius for the primary $M_1$\,=\,1.39\,$\pm$\,0.06\,\Msun and $R_1$\,=\,5.35\,$\pm$\,0.09\,\Rsun.
Using the asymptotic period spacing of the dipole mixed modes of $\Delta\Pi_1$\,=\,80.78\,s, \citetalias{Beck2018} determined the evolutionary state of the primary to be located on the RGB, which corresponds to the H-shell burning phase.
 Within the uncertainties, the global asteroseismic values of the primary from \citetalias{Beck2018} are in agreement with the recent values from the APOKASC-3 catalog \citep{APOKASC3}.

From time series spectroscopy, obtained with the High-Efficiency and High-Resolution Mercator Echelle Spectrograph \citep[HERMES,][]{Raskin2011}, \citetalias{Beck2018} found \KIC to be a double-lined spectroscopic binary (SB2). From disentangling the composite spectrum in Fourier space \citep{Ilijic2004}, \citetalias{Beck2018} found the mass ratio to be q\,=\,$M_1$/$M_2$\,=\,1.015\,$\pm$\,0.005. As expected for a binary born from the same cloud, the metallicity for the primary and secondary agree within the uncertainties ($[\mathrm{M/H}]_1$\,=\,-0.37\,$\pm$\,0.1\,dex and $[\mathrm{M/H}]_2$\,=\,-0.38\,$\pm$\,0.1\,dex, respectively; with a reference solar iron abundance of $Z_{\odot}$\,=\,0.0134). Despite their similarities in mass and metallicity, other fundamental parameters of the two components differ significantly, 
$\mathrm{T}_{\mathrm{eff},1}$\,=\,5020\,$\pm$\,100\,K vs. $\mathrm{T}_{\mathrm{eff},2}$\,=\,5650\,$\pm$\,70\,K, and 
$\log$\,$g_1$\,=\,3.14\,$\pm$\,0.2\,dex vs. $\log$\,$g_2$\,=\,3.48\,$\pm$\,0.3\,dex. From spectroscopy, \citetalias{Beck2018} determined that the primary and secondary contribute 60\%  and 40\% to the total flux, respectively. Due to the photometric dilution of the flux from each component, the relative flux variations of the oscillation modes are reduced by the same fraction. However, we do not expect the flux dilution to impact the extracted seismic parameters \citep[e.g. ][]{Sekaran2019}. Figure\,\ref{fig:HRD} depicts the positions of both components in the asteroseismic HRD.
  
 While the power excess of the secondary was found at 215\,$\pm$\,4\muHz, it was shown to be the reflection of the actual power excess at $\nu_{\mathrm{max},2}$\,=\,340\,$\pm$\,20\muHz, located above the Nyquist frequency of 283\muHz for the $\sim$30\,minute long-cadence data of \Kepler.
Because of the low signal-to-noise ratio (S/N) of the secondary's oscillation modes and the overlap with the primary's power excess, no value for the large frequency separation could be determined. From the position of the power excess, the He-core of the secondary is already degenerated. \citetalias{Beck2018} therefore considered the secondary an early RGB star too. Other criteria, based on the position of the secondary in the HRD shown in Fig.\,\ref{fig:HRD}, classifies the star as a late subgiant \citep[e.g. ][]{Diego2024}.

Another notable difference between the two components is their photospheric abundance of lithium. From the disentangled HERMES spectra, \citetalias{Beck2018} determined a lithium abundance for the primary of 1.31\,$\pm$\,0.08\,dex and the secondary 2.55\,$\pm$\,0.07\,dex, differing by a factor of 17. \citetalias{Beck2018} could only explain this observational difference, with both components being in a particular short-lived state of stellar evolution, the first dredge-up (FDU). During this brief phase of stellar evolution, located near the beginning of the RGB, the convective envelope deepens into the stellar interior. It causes the mixing of the material produced from hydrogen burning during the main sequence. This mixing causes a depletion of the lithium abundance, and a change in the carbon isotopes ratio and nitrogen abundance at the surface during the FDU \citep[][and references therein]{Roberts2024}.
Therefore, \citetalias{Beck2018} concluded that the secondary component of \KIC, which is richer in lithium than the primary, is in the early phase of the FDU. In contrast, the more massive primary evolved faster and is in a more advanced state of the FDU, causing a more effective mixing and lithium depletion.

For this paper, we reanalyzed the disentangled HERMES spectra assuming the same parameters as \citetalias{Beck2018} to determine the $\rm \alpha$-element abundances. The measurement has been made with the BACCHUS code \citep{Masseron2016} which includes MARCS model atmospheres \citep{Gustafsson2008} and linelists from \citet{Heiter2015}. This analysis indicated no enhancement nor depletion of either Mg, Si, or Ca, and provided an $\alpha$-element abundances of [$\alpha$/Fe]\,=\,0.0\,$\pm$\,0.1\,\dex for both components.

Since the analysis by \cite{Beck2014} and \citetalias{Beck2018}, the ESA \Gaia \citep{GaiaMain} mission has included \KIC in all three data releases. In the latest \Gaia DR3 \citep{GaiaDR3} \KIC has an apparent G mean combined magnitude of 9.6\,mag that have a color index $G_{BP}$\,-\,$G_{RP}$=\,0.60\,mag. The measured parallax is 2.23\,mas, corresponding to a target distance of about 445 pc. The system is not listed in the non-single star catalog of \Gaia DR3 \citep{Arenou2022}. The target renormalized unit weight error (\textit{ruwe}) value, an indicator for the object to be a single star (\textit{ruwe}\,$\sim$\,1) or potentially non-single star (\textit{ruwe}\,$>$\,1.4), is given as 1.051. \cite{BeckGrossmann2023} has shown that a significant percentage of known binaries (around 40 \% in their sample) exhibit a \textit{ruwe} below the generally accepted threshold of 1.4. Furthermore, neither multi-epoch photometry nor radial velocity was measured by \Gaia for this object.

A direct estimate of the distance of \KIC as the inverse of the parallax from DR3 ($\varpi$\,=\,2.23\,$\pm$\,0.01\,mas) results in a distance of d\,=\,448\,$\pm$\,5 pc. This is in agreement with the distances of $d=443^{+10}_{-6}$ pc, that is determined by the General Stellar Parametrizer from Photometry \citep[GSP-Phot,][]{Andrae2023}, which assumes a single star solution.  A distance modulus estimate from the asteroseismic luminosity and apparent magnitude values in \citetalias{Beck2018} results in a distance of $\sim$400 pc. However, it disagrees by more than 2-$\sigma$ with the multi-star classifier (MSC)\footnote{\url{https://gea.esac.esa.int/archive/documentation/GDR3/Data_analysis/chap_cu8par/sec_cu8par_apsis/ssec_cu8par_apsis_msc.html}} distance (d\,=\,$254^{+50}_{-114}$\,pc), which assumes an unresolved binary solution (as is the case for \KIC). \cite{GaiaDoc11} suggests that for some cases of unresolved binaries, the single-star solution of the GSP-Phot solution might be in better agreement with independent measurements compared to the MSC distance. The distance modulus solution and median photogeometric distance (d$_{\mathrm{med}}$\,=\,$445$\,pc) from \cite{Bailer2021Gaia} point in the same direction, so we decide to trust the value inferred from GSP-Phot of d\,=\,$443^{+10}_{-6}$\,pc as the reference for the distance to \KIC. 

Table\,\ref{tab:observables} summarizes all observables relevant for this work.

\section{Constructing the model grid} \label{sec:grid}

\begin{table}[t!]
\caption{Parameter range for model grid used in this work.}
\tabcolsep=6.5pt 

\centering
\begin{tabular}{lccc}
\hline
\hline
Parameter & & Value (range)  & Step size    \\ \hline
Mass, M                & [\Msun] & [1.20; 1.54]   & 0.01 \\

Initial Metallicity, $Z$ & $-$ & [0.004; 0.010]    & 0.001   \\
Initial helium, $Y$ & $-$ & [0.24; 0.30]& 0.01 \\
\hline
Mixing length, $\alpha_\mathrm{MLT}$ & $-$ & [1.4; 1.6] & 0.1 \\
Overshooting, $f_{\mathrm{ov}}$ & $-$  & 0.02 & $-$ \\ 

\hline            
\end{tabular}

\label{tab:parameters}

\tablefoot{The first two columns define the parameter and its unit used for the grid modeling. The third and fourth columns define the grid's value range and step size. The last two rows display the values used for mixing length $\alpha_\mathrm{MLT}$ and  Overshooting Parameter $f_{\mathrm{ov}}$ used in all modeling. }
\end{table}

\begin{table*}[t!]
\caption{Summary of cases considered for modeling \KIC and the associated constraints.}
\setlength{\tabcolsep}{10pt}
\renewcommand{\arraystretch}{1.5}
\centering
\begin{tabular}{cllll}

\hline
\hline
Case & & Observables used in figure of merit  & & Constraints in addition to Age$_{1}$\,=\,Age$_{2}$\\\hline
A0           &     \ldelim \} {4}{-10 mm}& 
\multirow{4}{1 mm}{$\mathrm{T_{eff,1}}$,\;$\mathrm{T_{eff,2}}$,\;$\log$\,$g_1$,\;$\log$\,$g_2$,\;$q$} & \rdelim \{ {4}{1 mm}&
$-$    \\
A1             &   & & &
$Z_1$\,=\,$Z_2$,\;$Y_1$\,=\,$Y_2$,\\ 
A2            &    & & &
$Z_{1,2}$\,=\,$\mathrm{Z_{spec}}$,\;$Y_1$\,=\,$Y_2$\\ 

A3           &    & & & 
$Z_{1,2}$\,=\,$\mathrm{Z_{spec}}$\,$\pm$\,$\sigma_{\mathrm{Z_{spec}}}$,\;$Y_1$\,=\,$Y_2$ \\ \hline
B1                &  \ldelim \} {3}{*}&\multirow{3}{*}{$\mathrm{T_{eff,1}}$,\;$\mathrm{T_{eff,2}}$\;$\log$\,$g_2$,\;$q$,\;$\nu_{\mathrm{max,1}}$} & \rdelim \{ {3}{1 mm}&
$Z_1$\,=\,$Z_2$, $Y_1$\,=\,$Y_2$ \\
B2                & & && $Z_{1,2}$\,=\,$\mathrm{Z_{spec}}$,~$Y_1$\,=\,$Y_2$ \\
B3                & &&
&
$Z_{1,2}$\,=\,$\mathrm{Z_{spec}}$\,$\pm$\,$\sigma_{\mathrm{Z_{spec}}}$,~$Y_1$\,=\,$Y_2$ \\ \hline
C1                &   \ldelim \} {3}{1 mm}&
\multirow{3}{*}{$\mathrm{T_{eff,1}}$,\;$\mathrm{T_{eff,2}}$,\;$\log$\,$g_2$,\;$q$,\,$\nu_{\mathrm{max,1}}$,\;$\Delta\nu_1$} & \rdelim \{ {3}{1 mm}&
$Z_1$\,=\,$Z_2$,~$Y_1$\,=\,$Y_2$ \\
C2                & &&
&
$Z_{1,2}$\,=\,$\mathrm{Z_{spec}}$,~$Y_1$\,=\,$Y_2$ \\
C3                & & &&
$Z_{1,2}$\,=\,$\mathrm{Z_{spec}}$\,$\pm$\,$\sigma_{\mathrm{Z_{spec}}}$,~$Y_1$\,=\,$Y_2$ \\ \hline
D1 &   \ldelim \} {3}{1 mm}&
\multirow{3}{*}{$\mathrm{T_{eff,1}}$,\;$\mathrm{T_{eff,2}}$,\;$\log$\,$g_2$,\;$q$,\;$\nu_{\mathrm{max,1}}$,\;$\Delta\nu_1$,\;$\Delta\mathrm{m_V}$} & \rdelim \{ {3}{1 mm}&
$Z_1$\,=\,$Z_2$,~$Y_1$\,=\,$Y_2$\\
D2 & & & &
$Z_{1,2}$\,=\,$\mathrm{Z_{spec}}$,~$Y_1$\,=\,$Y_2$\\
D3 & & & &
$Z_{1,2}$\,=\,$\mathrm{Z_{spec}}$\,$\pm$\,$\sigma_{\mathrm{Z_{spec}}}$,~$Y_1$\,=\,$Y_2$\\ \hline

\hline            
\end{tabular}

\label{tab:cases}

\tablefoot{The first column names the different optimization cases. The second column indicates those parameters included in the \chisq optimization. The third column indicates any constraints in the model input, in addition to the requirement of both components exhibiting the same age, which has been applied to all cases.}
\end{table*} 

To model \KIC, we used the 1D-stellar evolutionary code MESA \citep[Modules for Experiments in Stellar Astrophysics;][version r23.05.1]{Paxton2011,Paxton2013,Paxton2015,Paxton2018,Paxton2019,Jermyn2023} and the stellar oscillations code GYRE \citep[][version 7.1]{Townsend2013} in a combined grid modeling approach. Using MESA, we calculated a grid of single-star models to search for the best combination of models for the primary and secondary under the constraints imposed by the common age (Age$_{1}$\,=\,Age$_{2}$) and chemical composition. For the grid of models, we varied the stellar mass M, metallicity $Z$, and helium abundances $Y$ in the ranges and stepsizes given in Table\,\ref{tab:parameters}. Regarding the heavy element distribution adopted in the stellar modeling, we rely on a solar-scaled mixture \citep{GrevesseSauval1998}. This choice is fully supported by the reanalysis of
the spectra (Sect.\,\ref{sec:observations}) that shows that our stellar target does not show any evidence of $\alpha-$element enhancement.   

For the stellar mass range of the grid, we chose an interval centered on the mass of the primary reported by \citetalias{Beck2018} from an asteroseismic analysis and extended three times the reported uncertainty towards higher and lower masses.
This resulted in a lower limit of 1.22\,\Msun and an upper limit of 1.54\,\Msun. To account for the lower mass of the secondary of 1.37\,\Msun, we set the lower limit to 1.20\,\Msun. With $\sim$1\% being the lowest feasible observational uncertainty for stellar mass, we choose a stepsize of 0.01\,\Msun.

    We followed a similar approach for the initial metallicity, creating a grid range of two times the uncertainty centered on the spectroscopic value of the metallicity. This resulted in a range of -0.57\,dex to -0.17\,dex. Prior to modeling the with MESA, we converted the spectroscopic metallicity into the fractional mass component using the following equation:
\begin{equation}
    Z_{\star} \sim Z_{\odot} \times 10^{[M/H]}\,{.}
    \label{eq:easyfinalmetal}
\end{equation}
When assuming the solar iron abundance value used for the spectroscopic analysis in \citetalias{Beck2018} ($Z_{\odot}$\,=\,0.0134), the heavy element abundance of the primary of ${[}$M/H${]}$\,=\,-0.37\,dex translates into $Z_{\mathrm{spec}}$\,=\,0.006. Correspondingly, the upper and lower boundaries of our modeling result in $Z_{\mathrm{low}}$\,=\,0.004 and $Z_{\mathrm{high}}$\,=\,0.010. As a stepsize, we decided to use 0.001, as translated in logarithmic space, this is approximately the order of the lowest feasible observational uncertainty.

For the initial helium abundance, which was added as a free parameter to account for a potential mass-initial helium degeneracy \citep[e.g.][and references therein]{Valle2014,Verma2022Helium}, we chose a range similar to the one used in \cite{Lebreton2014helium}, who did a similar modeling for a single star. Our range starts just below the primordial helium abundance \citep[$Y_{prim}$\,=\,0.245;][]{Peimbert2007He} and goes up to a value of 0.30, resulting in a range of 0.24\,$\leq$\,Y$_\mathrm{init}$\,$\leq$\,0.30, with a stepsize of 0.01. 

Apart from these physical constraints, modeling stars also requires attention to the use of adequately calibrated parameters describing the microscopic and macroscopic mixing process of the physics acting in the stellar interior. One of the most important transport mechanisms is convection, which in 1D models is typically treated by mixing length theory \citep[MLT,][]{Paxton2011}. In this notation, $\alpha_{\mathrm{MLT}}$ describes the typical distance in pressure scale heights a convective cell is transported before it dissolves \citep[for a detailed review on MLT see][and references therein]{JoyceTayar2023}.
The mixing length parameter is difficult to determine and is model-dependent, but it has large effects on the position of the RGB. 

With 1.4\,\Msun and sub-solar metallicity, the main sequence progenitors of the two components of \KIC were mid-F-type dwarfs. By modeling the measured lithium abundances \citetalias{Beck2018} showed that the progenitor stars must have been rigidly rotating and had weak macroscopic mixing processes. This behavior indicates that the progenitors did not develop significant convective envelopes during their H-core-burning phase that would have reduced the Li abundance \citep[see][and references therein]{Beck2017}. Convection would have become increasingly important as both components develop extensive convective envelopes and evolved away from the main sequence towards the FDU event.
As a starting value, we chose the solar-calibrated value of $\alpha_{\mathrm{MLT}}$\,=\,2.0 used as a default in MESA \citep{Paxton2011}. As the literature suggests lower than solar mixing length for similar stars to our targets \citep[e.g.][]{Trampedach2014ML}, we tested different values as described in Appendix\,\ref{sec:alphamlt}. From the results of this test, we decided to use $\alpha_{\mathrm{MLT}}$ between 1.4 and 1.6 in the remainder of the paper. As the models with free $\alpha_{\mathrm{MLT}}$ within the three allowed values (1.4, 1.5 and 1.6) showed a better fit with the observables compared to a forced equal $\alpha_{\mathrm{MLT}}$, we decided not to enforce strict mixing length equality for both components.

The overshooting (or undershooting) of convective cells into convectively stable radiative regions influences the diffusive mixing in convective layers of the star and hence needs to be accounted for in modeling. While the progenitor stars had insignificant convective envelopes, a certain impact of mixing length on the main-sequence evolution is expected due to the small convective core in these stars. The effect grows with the convective envelopes as the two stars rise on the sub-giant branch (SGB) and RGB, respectively. As widely used in literature \citep[e.g. ][and references therein]{Constantino2015over}, we treat overshooting as exponential for our modeling. We tested the impact of different values for the overshooting parameter on our parameter determination by calculating the model grid for a sample case (Case C) for three different values of $f_{\mathrm{ov}}$: 0.020, 0.014, 0.002 and no overshooting overall. Because the changes in the model parameters with different overshooting parameters throughout all subcases were at least an order of magnitude smaller than the changes induced by any other parameter in the model grid, we decided to use a fixed value for the overshooting for the remainder of this work.  We decided to use a value of 0.02 because this best reproduces the observables of the system.

Another parameter requiring attention when performing 1D-stellar modeling is the (atmospheric) boundary conditions, particularly the temperature (and pressure) at a given optical depth $\tau$. In MESA, this is usually given as the $T_{\tau}$ relation, set default to Eddington grey relation \citep{Paxton2011}. As 
extensively discussed by \citet{Salaris2018} and, more recently, by \citet{Creevey2024}, the choice of the $T_{\tau}$ relation in combination with the choice of $\alpha_{\mathrm{MLT}}$ can have a significant impact on the effective temperatures on the RGB. We tested different implementations of the $T_{\tau}$ relation but decided to keep the Eddington grey relation because, in our case, it provided a better agreement with observations. 

Using GYRE, we calculated the radial modes in the frequency range of the power excess for each model, ensuring all central frequencies along every evolutionary step up to the luminosity limit are encompassed. An example of the GYRE input file used for the calculation is shown in Appendix\,\ref{app:GYRE}. We derived a local \dnu by using the central radial modes and their radial orders from Table\,7 in \citetalias{Beck2018}, and applying a linear regression to them, which resulted in a revised \dnu\,=\,12.89\,$\pm$\,0.02\,\muHz. This value is slightly larger than the value reported for the local large frequency separation from auto-correlation in \citetalias{Beck2018}. To correct the \dnu from GYRE for surface effects and the deviation from the asymptotic approximation, we used the empirically determined correction factor $f_{\Delta\nu}$ introduced in Eq.\,16 of \cite{Li2023Surface} and applied it to the \dnu acquired from the GYRE frequencies. We note that individual oscillation modes observed in active stars can experience higher order shifts in addition to surface effects and deviation from asymptotic regime \citep{Salabert2016}. We bypass this uncertainty by using the local large-frequency separation \dnu, which will be less impacted by potential shifts, instead of individual modes. To be consistent with the definition of \dnu derived from the GYRE oscillations, we recalculated the observed \dnu similarly.

 We set MESA to calculate the models with twice default temporal resolution (\texttt{time\_delta\_coeff}\,=\,0.5). 
Further modifications of the timesteps in MESA were necessary to model \dnu with sufficient resolution compared to the uncertainty from observation (0.03\,$\mu\mathrm{Hz}$). Therefore, we had to set a maximum timestep of 0.2\,Myr for all models cooler than 6000\,K on the SGB and RGB and reduce it to 0.1\,Myr for all masses exceeding 1.44\,\Msun. Those GYRE-related timestep modifications are described in more detail in Sect.\,\ref{app:resolution}. 

Following the approach in the Python package \texttt{multimesa}\footnote{available at \hyperlink{https://github.com/rjfarmer/multimesa}{https://github.com/rjfarmer/multimesa}}, the inlists were split up into a MESA Base inlist, constant for all computations, and another inlist file including only the parameters varied between the models (Table\,\ref{tab:parameters}). For the calculations of the model grid, we built the inlists based on the \texttt{1M\_pre\_ms\_to\_wd} test suite, adapting it with the criteria mentioned above. An example of the MESA inlist, compiled from both the MESA Base and variable inlist used for our modeling, is provided in Appendix \ref{app:MESA}.

The resulting model grid encompasses 12\,096 single-star models. For higher computational efficiency, we evolved the models to a luminosity limit of  $\log L/L_{\odot}$\,=\,1.5 as a stopping criterion for the model calculations. This value is well above the observationally determined luminosity of the primary of $\log L/L_{\odot}$\,$\sim$\,1.2.

\section{Figure of merit and error estimation}
\label{sec:bestfit}
This section will define a primary figure of merit used to determine the best-fitting models compared to the observations. Furthermore, we will discuss how the errors in fitting the observations are determined.

\subsection{Definition of the figure of merit} 
\label{sec:figure}
To quantify the best combination of models for the primary and secondary from the model grid generated in Sect.\,\ref{sec:grid}, we compare the modeled parameters for each timestep to the spectroscopic and seismic observables (Sect.\,\ref{sec:observations}). 
We use a reduced \chisq-minimization approach, with its general form given by, 
\begin{equation}
    \chi^2_\mathrm{red} = \frac{1}{n-m}\sum_{n=1}^{N} \left( \frac{(y_{n,\text{obs}}-y_{n,\text{mod}})}{\sigma_n} \right)^2\,,
    \label{eq:chitot}
\end{equation}
for a model with $N$ parameters, where $y_{n,\text{obs}}$ represents the observed quantity  and $ y_{n,\text{mod}}$ the respective quantity from modelling, with $\sigma_n$ being the observational error and $n$ being the number of observables and $m$ the number of fitted parameters. For every model combination, we determine the timestep of the lowest \chisq, referred to as  $\chi^2_\mathrm{min}$, and compare those minima with each other in our further analysis.

The dynamical choice of timesteps between consecutive models in MESA, as described in \cite{Paxton2011}, causes the timesteps to be non-uniform. We interpolate the steps in the time domain for every relevant parameter for each given model track to compare two models at identical times before performing the \chisq calculation. This ensures that both components have the same age (Age$_{1}$\,=\,Age$_{2}$).  We rebinned the secondary's evolutionary tracks to the primary's timesteps. Models younger than 0.5\,Gyr are removed to prevent false minima in the \chisq calculation when the star passes close to the observed values on the HRD during the pre-main-sequence phase. The frequency of maximum oscillation is calculated from the model's position on the HRD using the scaling relations \citep{Kjeldsen1995a}.

For the \chisq calculation, we allowed every combination of the models (representing primary and secondary, respectively) that fulfilled the criteria that the mass of the primary was equal or larger than the mass of the secondary ($M_2$\,$\leq$\,$M_1$), due to the solid constraint of the spectroscopic mass ratio $q$\,=\,$M_1$/$M_2$\,=\,1.015\,$\pm$\,0.005.

\subsection{Error estimation}
\label{sec:error}

As outlined in Sect.\,\ref{sec:introduction}, error estimation is crucial in stellar age determination. Equation 2 considers the observables' reported uncertainties (Sect.\,\ref{sec:observations}). The estimation of errors from the minimization in the $ \chi^{2}$ statistics for the fitted parameter $\sigma_{i,y}$ is given as \citep{bevington2003Error},
\begin{equation}
    \sigma_{i,y}= \sigma_n \chi^{2},
\end{equation}
with $\sigma_n$ being the statistical error from the respective observable. If the smallest \chisq is above one, we assume the observational uncertainty as an error for our analysis because lower (formal) uncertainties do not have a physical meaning in this context. This approach, however, does not appropriately work for age estimation since there is no trivial mathematical description connecting the input parameters with the respective age of the system.

For the error determination in age from the calculated \chisq, we evaluate all \chisq\,$<$\,1, which, from a statistical viewpoint, have the same significance. From the scatter in age for these values with the same significance, we determine the edges of the area of confidence and give them as the upper and lower boundaries for the 1-$\sigma$ uncertainty level. In case singular model points occur at a significantly higher or lower age with \chisq\,$<$\,1, we test if those are consistent with stellar models for the observational mass/metallicity, and if not, remove them as outliers from the analysis.

Besides statistical errors, we are aware that the fitting procedure itself can introduce biases into the results of the modeling \citep[e.g.][]{Valle2014,Valle2015a}. Therefore, we applied a consistency check by recovering the secondary’s parameters from the primary’s HRD position and age using two model tracks with the same initial parameters, only differing in the mass by the mass ratio $q$. We were successfully able to recover the secondary's HRD position within the margins of the 1-$\sigma$ uncertainties of the observables.

\section{Determination of the best fitting models for different sets of parameters}
\label{sec:cases}

Using the figure of merit defined in Sect.\,\ref{sec:figure}, in this section, we test different parameter combinations for calculating \chisq.  In the following subsections, we discuss the results of several cases where different sets of observables and constraints are used to find the best-fit model. The cases are summarized in Table\,\ref{tab:cases}. To show the improvements from strongly physically motivated parameter combinations, we also provide cases as a reference point that are less physically motivated but commonly used if fewer observational constraints are available.

\subsection{Case A: Modeling of spectroscopic constraints} \label{sec:spec}

Using the \chisq figure of merit introduced in its general form in Eq.\,\ref{eq:chitot}, we first define the 'spectroscopic case'. The parameters used are the effective temperatures, $T_{\text{eff,1}}$ and $ T_{\text{eff,2}}$, the surface gravities $\log g_1$ and $\log g_2 $ of the primary and secondary component, respectively as well as the mass ratio $q$ from spectroscopy. As we used the metallicity as an input parameter of the grid, we did not include it in the figure of merit, to avoid a circular argument. This parameter combination results in the formulation of the \chisq estimation for case A to be,
\begin{equation}
\label{eq:chispec}
    \chi_{A}^{2} = \chi_{T_{\text{eff,1}}}^{2} + \chi_{T_{\text{eff,2}}}^{2}+ \chi_{\log g_1}^{2} + \chi_{\log g_2}^{2} + \chi_{q}^{2},
\end{equation}
where each \chisq is calculated according to the term after the sum in  Eq.\,\ref{eq:chitot}.

To exploit the potential provided by the co-evolution of the two components of the binary system, we subsequently apply stronger constraints justified by their stellar binarity and evolutionary history. We generate four different cases for each formulation of the figure of merit. In the simplest case, A0, the residuals according to Equation\,\ref{eq:chispec} are calculated for all the combinations of models that satisfy the aforementioned mass constraint, M$_2$\,$\leq$\,M$_1$, independent of the primary's or secondary's model metallicity.
\begin{figure}[t!]
    \includegraphics[width=\columnwidth
    ]{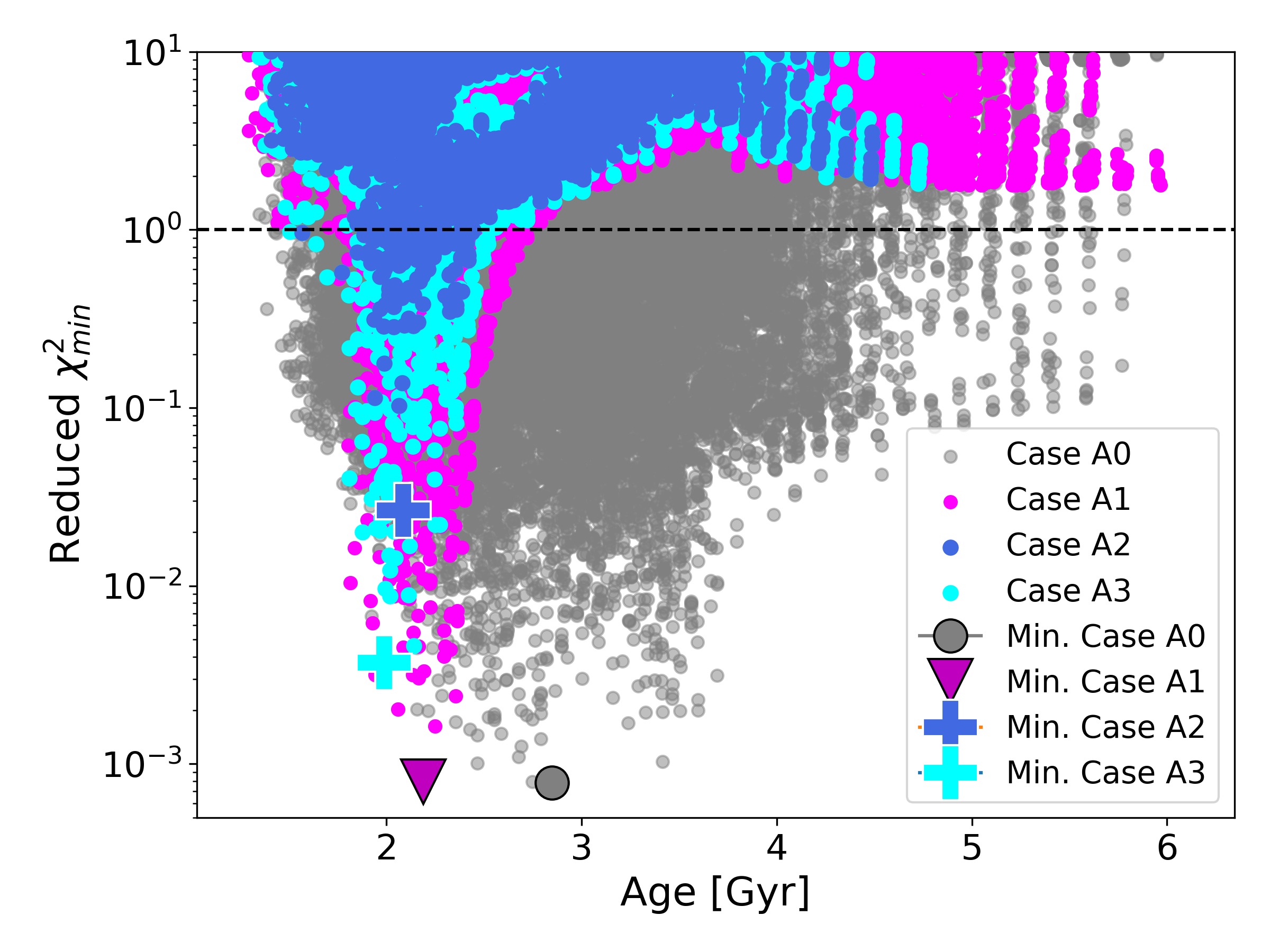}
    \caption{Comparison of the best fitting solutions for case A regarding the treatment of metallicity. The grey solutions have the metallicity varying freely in the modeled range (Case A0), the magenta solution only allows values where the metallicity of the primary equals the metallicity of the secondary (Case A1), while the royal blue solutions force both metallicities to be equal to the spectroscopic solution (Case A2). The cyan part only allows values in the range of $1\sigma$ around the spectroscopic value (Case A3). The magenta triangle, blue cross, and grey dot mark the respective minima of the corresponding datasets.
    }
    \label{fig:chi_comp}
\end{figure}
\begin{figure}[t!]
    \includegraphics[height=0.7\columnwidth,width=\columnwidth
    ]{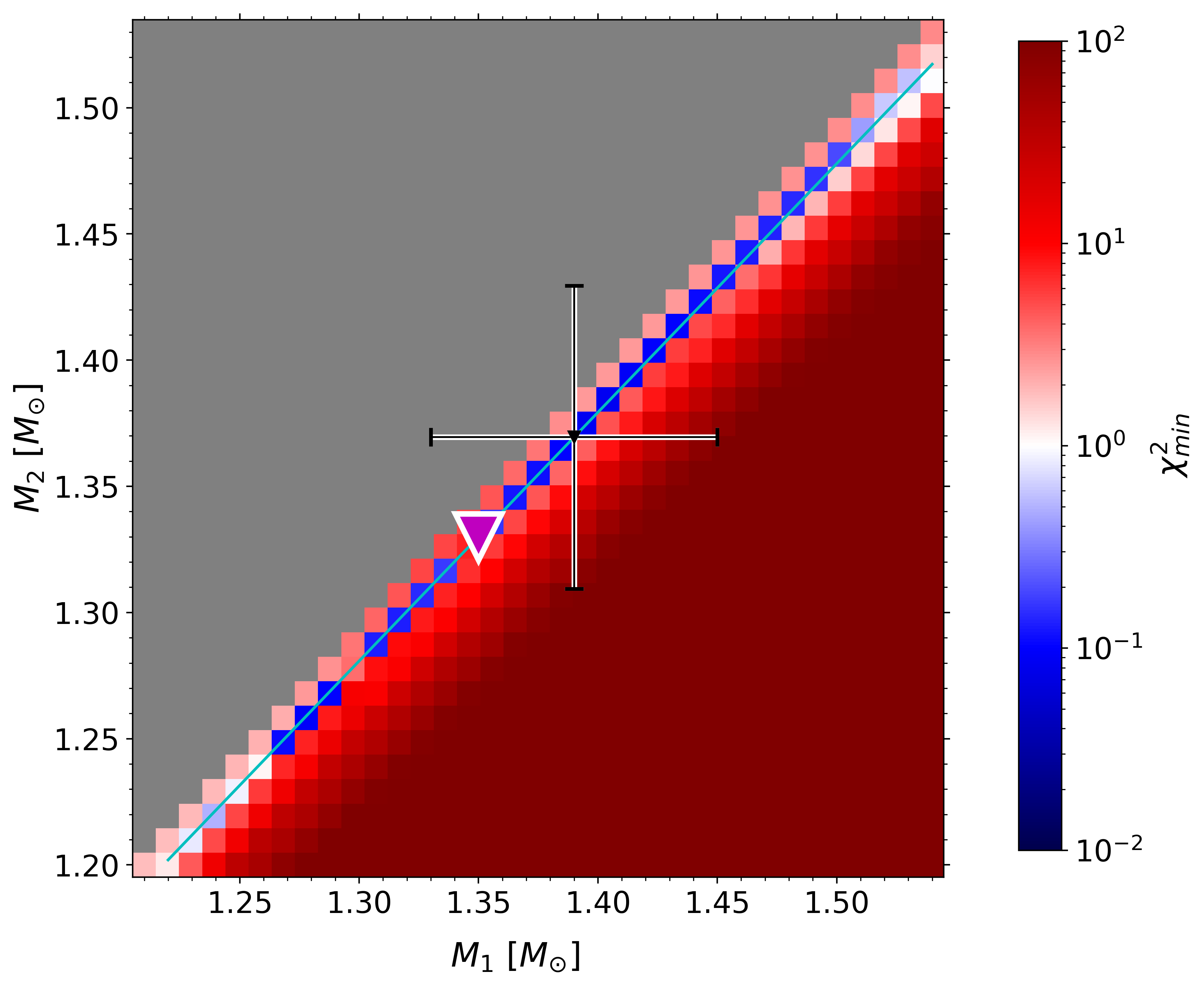}
    \caption{Secondary mass and \chisq-minima as a function of primary mass from the models with the condition of equal metallicity of both components $Z_1$\,=\,$Z_2$ (Case A1). The color bar indicates the respective \chisq-minima on a logarithmic scale. The residuals were calculated with enforcing $M_1$\,$\geq$\,$M_2$; the gray area represents the respective space where no \chisq is calculated due to this condition. The seismic solution, including its error bars from \citetalias{Beck2018}, is given in black, the absolute \chisq-minimum is given in magenta triangle, and the green line represents the spectroscopic mass ratio of $q$\,=\,1.015.}
    \label{fig:masscomparison}
\end{figure}
\begin{figure}[t!]
\includegraphics[width=\columnwidth
]{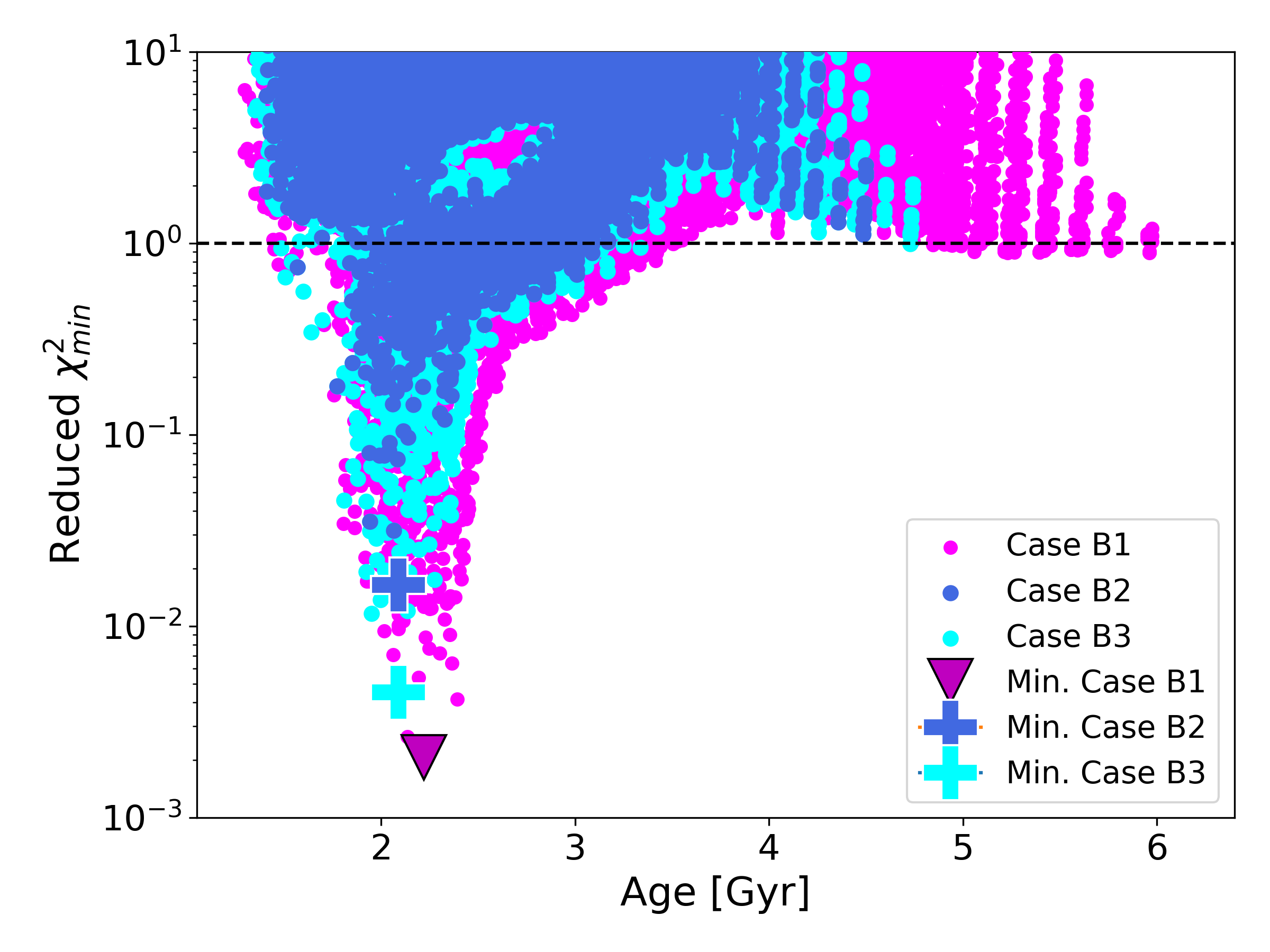}
\caption{Comparison of the best fitting solutions regarding the treatment of metallicity (set equal for primary and secondary: magenta; set equal and to spectroscopic value: royal blue; set equal and within a 1-$\sigma$ range of the spectroscopic metallicity: cyan) for the case of including \num in the figure of merit (Case B).  \hbox{Symbols and colors are identical to Fig.\,\ref{fig:chi_comp}.}}
    
\label{fig:chi_dnu}
\end{figure}
Given that both stars were born from the same cloud, we apply constraints on the metallicity. For simplicity, this work assumes the initial heavy element abundance equaling the current heavy element abundance. We tested if the assumption of $Z$\,=\,$Z_\mathrm{ini}$ causes deviations in our parameter determination but received a null result since the theoretical difference between the initial and current metallicity is at least one order of magnitude smaller than the uncertainty from the spectroscopic observations \citep{Salaris:Casissi,Nissen2018}.
For the case A1, we require the metallicity of both components to be equal (binary condition, $Z_1$\,=\,$Z_2$). 
In the case of A2, we enforce the same condition but additionally require the metallicity to be equal to the spectroscopic value ($Z_{1,2}$\,=\,$Z_{\mathrm{spec}}$). For case A3, we allow metallicities to vary within their 1-$\sigma$ uncertainty around the spectroscopic value ($Z_{1,2}$\,=\,$Z_{\mathrm{spec}}$\,$\pm$\,$\sigma_{Z_{\mathrm{spec}}}$). For all three cases A1-A3, we also force the initial helium to be identical for the primary and secondary (Y$_1$\,=\,Y$_2$).
    
We compare the \chisq distribution as a function of the age of cases A0-A3 in Fig.\,\ref{fig:chi_comp}.
The age of the system for the different cases is taken from the mean of all \chisq values present below 1, with 
the errors estimated as described in Sect.\,\ref{sec:error}. The least constrained case A0 provides an age of $2.68^{+3.12}_{-1.30}$\,Gyr, the three cases with increasing constraints, A1, A2, and A3 lead to ages of $2.19^{+0.53}_{-0.69}$\,Gyr, 2.08\,$^{+0.36}_{-0.53}$\,Gyr and $2.09^{+0.35}_{-0.52}$\,Gyr, respectively. Additionally, the minimum \chisq increases by about an order of magnitude towards \chisq\,=\,1 for each case (A0 to A3).

Figure\,\ref{fig:masscomparison} shows the \chisq in the plane of the masses of the primary and secondary components for case A1 and the associated best-fitting solution. A clear trend following a diagonal line is visible, corresponding to the spectroscopic mass ratio q\,=\,1.015. For case A0, A1, and A3 we observe the best fitting solution at masses of $M_1$\,=\,1.35\,$\pm$\,0.06\,\Msun and $M_2$\,=\,1.33\,$\pm$\,0.06\,\Msun. Only for case A2 do we observe slightly higher masses of $M_1$\,=\,1.36\,$\pm$\,0.06\,\Msun and $M_2$\,=\,1.34\,$\pm$\,0.06\,\Msun.

\begin{figure}[t]
    \centering
        \includegraphics[width=\columnwidth
    ]{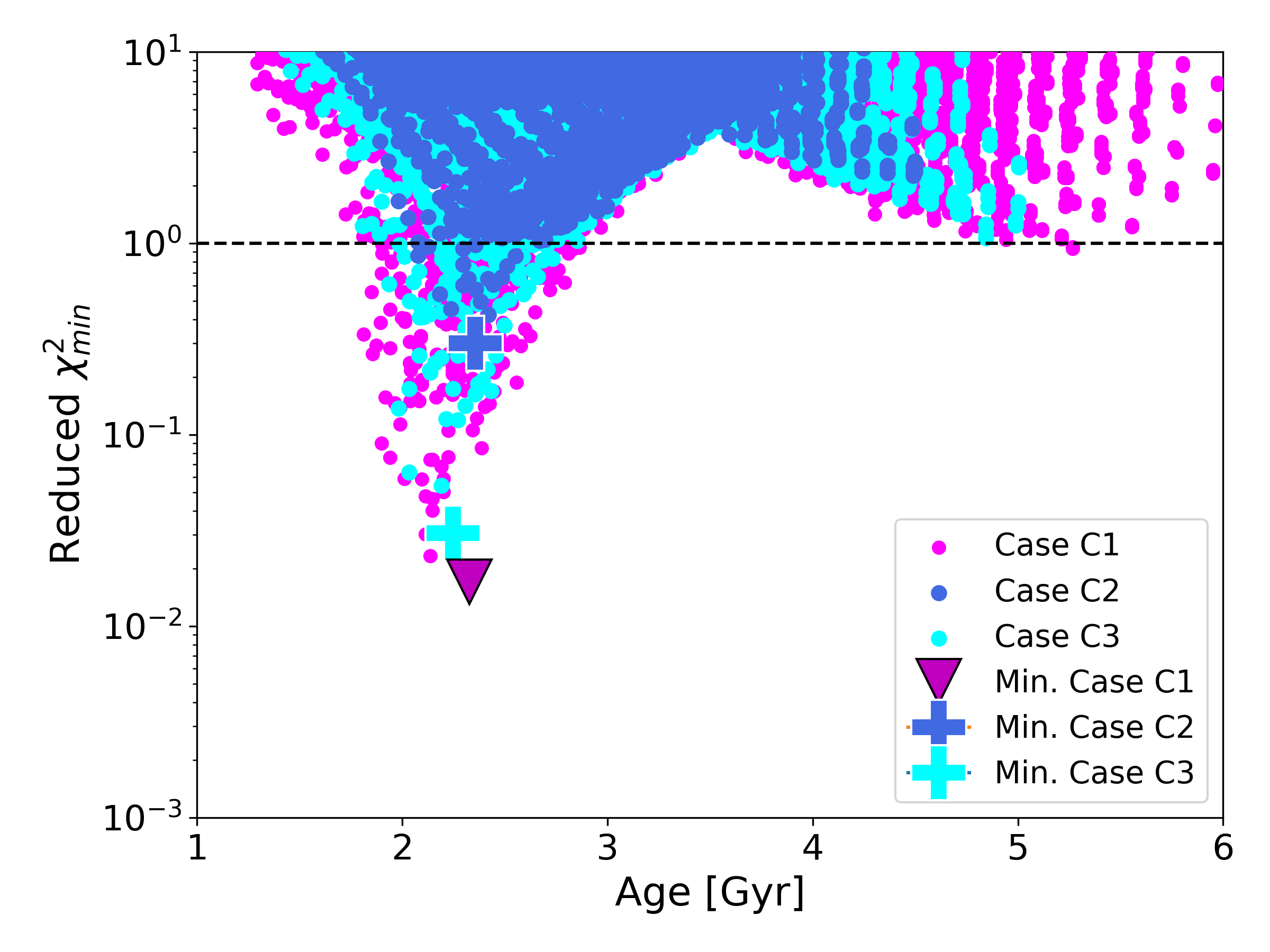}
    \caption{Comparison of the best fitting solutions regarding the treatment of metallicity (set equal for primary and secondary: magenta; set equal and to spectroscopic value: royal blue; set equal and within a 1-$\sigma$ range of the spectroscopic metallicity: cyan). The figure of merit used includes the local \dnu and \num of the primary in the figure of merit (Case C). Symbols and colors are identical to Fig.\,\ref{fig:chi_comp}. The dashed horizontal line shows where \chisq\,=\,1.}
    \label{fig:chi_comp_numax}
    
\vspace{5mm}
    \centering
    \includegraphics[height=0.75\columnwidth,width=\columnwidth
    ]{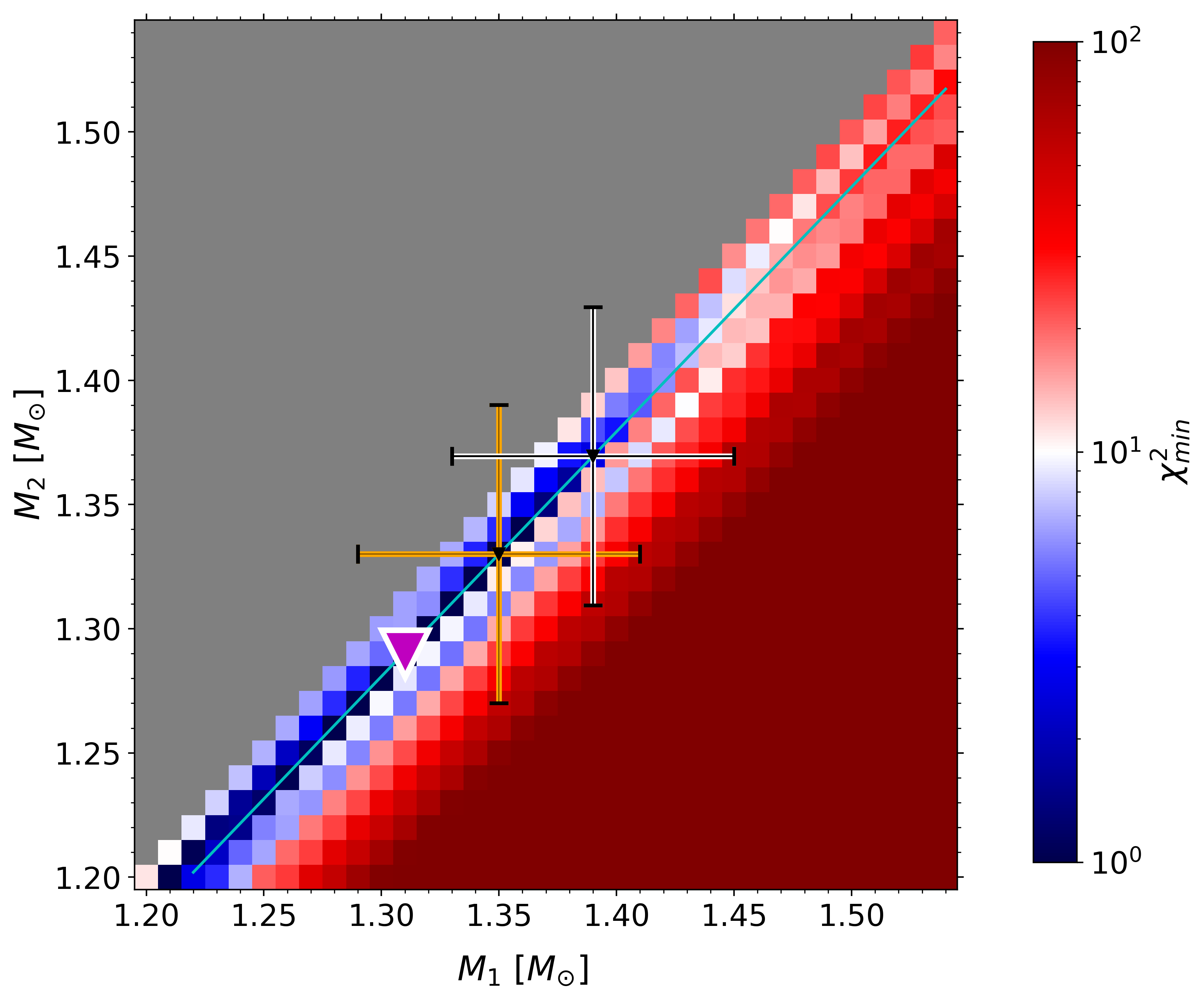}
    \caption{The residuals of the \chisq-Minimization as a function of the primary mass $M_1$ and secondary mass $M_2$ with the condition of equal metallicity within the uncertainty range of the spectroscopic value of both components $Z_{1,2}$\,=\,$\mathrm{Z_{spec}}$\,$\pm$\,$\sigma_{\mathrm{Z_{spec}}}$ (Case C3); with the inclusion of \dnu in the figure of merit. Colors and symbols are the same as in Fig.\,\ref{fig:masscomparison}. We added the revised seismic masses obtained from the corrected \dnu value as an orange cross.}
    \label{fig:dnumass}
\end{figure}

\begin{figure}[t]
    \centering
    \includegraphics[width=\columnwidth
    ]{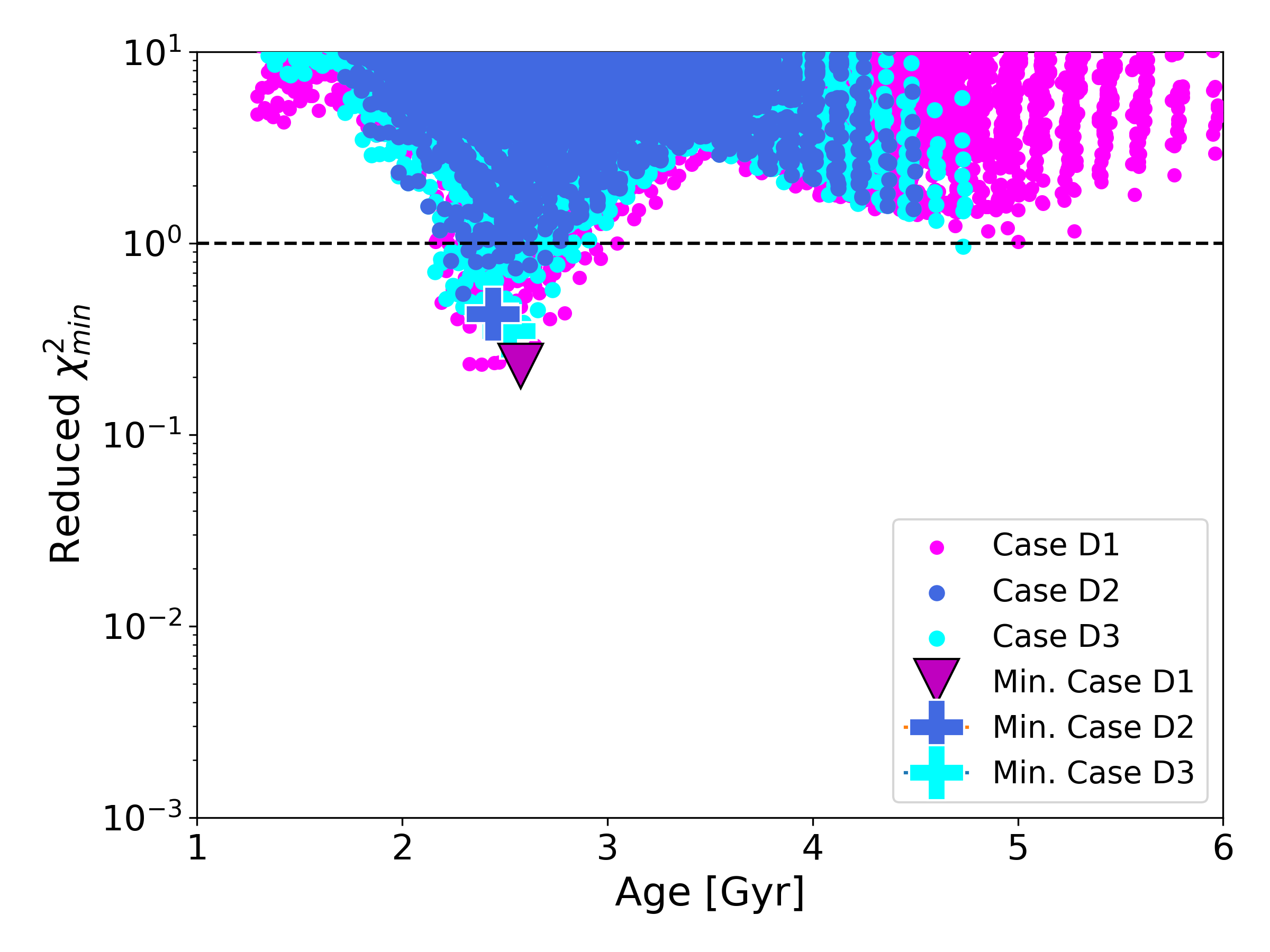}
    \caption{Comparison of the best fitting solutions regarding the treatment of metallicity (set equal for primary and secondary: magenta; set equal and to spectroscopic value: royal blue; set equal and within a 1-$\sigma$ range of the spectroscopic metallicity: cyan) for the case of including the local \dnu of the primary and the magnitude difference of both components into the figure of merit (Case D). Symbols and colors are identical to Fig.\,\ref{fig:chi_comp}. The dashed horizontal line shows where \chisq\,=\,1.}
    
    \label{fig:agenumax2}
\vspace{5mm}
    \centering
    \includegraphics[height=0.75\columnwidth,width=\columnwidth
    ]{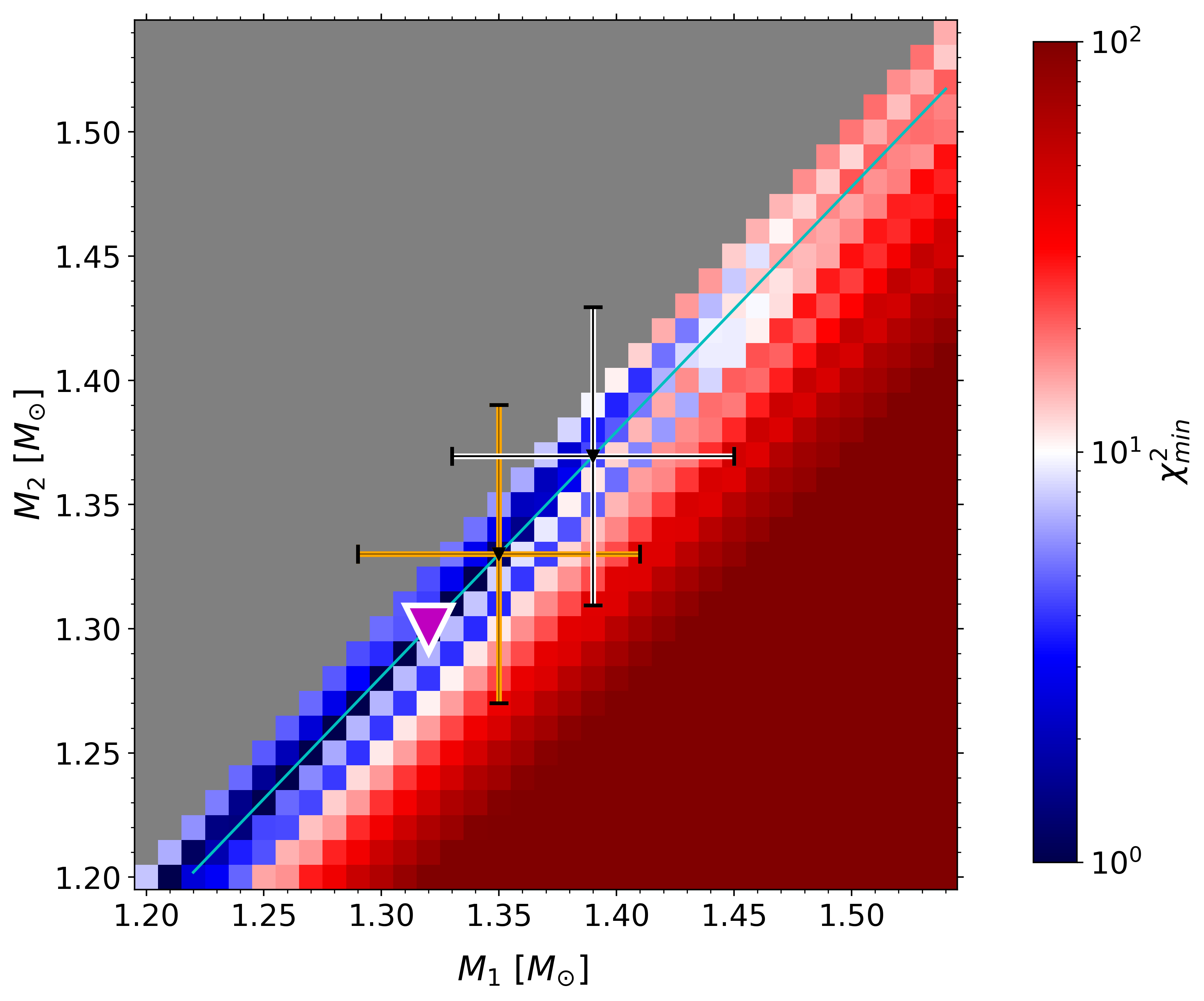}
    \caption{The residuals of the \chisq-Minimization as a function of the primary mass $M_1$ and secondary mass $M_2$ with the condition of equal metallicity within the uncertainty range of the spectroscopic value of both components $Z_{1,2}$\,=\,$\mathrm{Z_{spec}}$\,$\pm$\,$\sigma_{\mathrm{Z_{spec}}}$ (Case D3); with the inclusion of \dnu and the magnitude difference $\Delta m$ in the figure of merit. Colors and symbols are the same as in Fig.\,\ref{fig:masscomparison}. We added the revised seismic masses obtained from the corrected \dnu value as an orange cross.}
    \label{fig:dnumassD}
\end{figure}

\subsection{Case B: Modeling including \num -constraint} \label{sec:seism}

To evaluate the surplus of additional asteroseismic constraints provided by the stellar oscillations, we now introduce the global seismic parameters \num the figure of merit (case B). The corresponding value of the primary's \num of the stellar models is obtained with the scaling relations \citep{Kjeldsen1995a}. Because \num is strongly correlated with the surface gravity g via the scaling relations \citep{Brown1991,Kjeldsen1995a,Kallinger2014},
\begin{equation}
\nu_{\mathrm{max}} \propto \nu_{\mathrm{ac}} \propto g T_{\mathrm{eff}}^{-1/2}, 
\label{eq:numaxlogg}
\end{equation}
we remove the latter as an observable. Therefore, the following formulation of the figure of merit for case B is, 
\begin{equation}
\label{eq:chitot3}
    \chi_{B}^{2} = \chi_{T_{\text{eff,1}}}^{2} + \chi_{T_{\text{eff,2}}}^{2}+ \chi_{\log g_2}^{2} + \chi_{q}^{2} +  \chi_{\nu_{\mathrm{max,1}}}^{2}. 
\end{equation}

For case B, the minima of the \chisq for the model combination in the age-\chisq plane is given in Fig.\,\ref{fig:chi_dnu}. The cases B1, B2, and B3 lead to ages between 2.09 and 2.22\,Gyr,
with the exact values in Table\,\ref{tab:results}. While the minima for case B1 ($Z_1$\,=\,$Z_2$) is prominent, the minima for B2 ($Z_{1,2}$\,=\,$\mathrm{Z_{spec}}$) and B3 ($Z_{1,2}$\,=\,$\mathrm{Z_{spec}}\,\pm\,\sigma$) are more spread out, suggesting that constraints beyond binarity could lead to an over-fitted solution.
Regarding the mass, the best fitting model suggests a primary mass of $M_1$\,=\,1.37\,$\pm$\,0.06\,\Msun, and secondary mass of $M_2$\,=\,1.35\,$\pm$\,0.06\,\Msun, located at the upper end of the 1-$\sigma$ errorbars from observation. All cases B1-B3 suggest a spectroscopic metallicity of $Z_{1,2}= 0.006$\,$\pm$\,0.002 and a helium abundance of $Y_{1,2}= 0.30$\,$\pm$\,0.03.

\subsection{Case C: Modeling including \dnu -constrain} \label{sec:dnu}

As a next step, we include \dnu of the primary in the figure of merit. Therefore, the figure of merit for our case C reads,
\begin{equation}
\label{eq:chitot4}
    \chi_{C}^{2} = \chi_{T_{\text{eff,1}}}^{2} + \chi_{T_{\text{eff,2}}}^{2} + \chi_{\log g_2}^{2} + \chi_{q}^{2} + \chi_{\nu_{\mathrm{max,1}}}^{2} +\chi_{\Delta\nu_1}^{2}
\end{equation}
The addition of \dnu in the figure of merit should bring the advantage of being independent of scaling relations but instead rely on observed frequencies. The large frequency separation is calculated using GYRE as outlined in Sect.\,\ref{sec:grid}.
Again, the numbering of the cases C1 to C3 corresponds to the same constraints on mass, metallicity, and initial helium as described for the cases of the groups A and B and are listed in Table\,\ref{tab:cases}.

Using this new set of observables, the best-fit models have ages in the range of between 2.25\,Gyr and 2.35\,Gyr (Tab.\,\ref{tab:results}), with the variation between the cases with the different initial metallicity constraints, C1 to C3, being less than 0.1\,Gyr (Fig.\,\ref{fig:chi_comp_numax}). Regarding the mass, the best-fitting models point to a lower mass than in previous cases, down to 
$M_1$\,=\,1.31\,$\pm$\,0.06\,\Msun and $M_2$\,=\,1.29\,$\pm$\,0.06\,\Msun, in case of cases C2 and C3, for the primary and secondary mass, respectively (Fig.\,\ref{fig:dnumass}).

\begin{table*}[t!]
\caption{Summary of cases considered for modeling \KIC and the associated stellar parameters determined from \chisq minimization.}
\tabcolsep = 10pt 
\renewcommand{\arraystretch}{1.4} 
\centering
\begin{tabular}{ccccccccc}
\hline
\hline

Case & $M_1$               & $M_2$               & $Y$                   & $Z$                    & $\alpha_{\mathrm{MLT,1}}$ & $\alpha_{\mathrm{MLT,2}}$ & Common Age                      & \chisq \\
     & [\Msun]             & [\Msun]             &                     &                      &       &       & [Gyr]                    &        \\  \hline
A0   & 1.35\,$\pm$\,0.06\, & 1.33\,$\pm$\,0.06\, & 0.28\,$\pm$\,0.03\, & 0.004\,$\pm$\,0.002\, & 1.5   & 1.5   & 2.68\,$^{+3.12}_{-1.30}$   & 0.001  \\
A1   & 1.35\,$\pm$\,0.06\, & 1.33\,$\pm$\,0.06\, & 0.28\,$\pm$\,0.03\, & 0.004\,$\pm$\,0.002\, & 1.5   & 1.5   & 2.19\,$^{+0.53}_{-0.69}$ & 0.001    \\
A2   & 1.36\,$\pm$\,0.06\, & 1.34\,$\pm$\,0.06\,  & 0.28\,$\pm$\,0.03\, & 0.006*                & 1.4   & 1.6   & 2.08\,$^{+0.36}_{-0.53}$ & 0.05    \\
A3   & 1.36\,$\pm$\,0.06\, & 1.34\,$\pm$\,0.06\, & 0.30\,$\pm$\,0.03\, & 0.006\,$\pm$\,0.002\, & 1.4   & 1.5   & 1.99\,$^{+0.32}_{-0.41}$ & 0.008    \\  \hline
B1   & 1.37\,$\pm$\,0.06\, & 1.35\,$\pm$\,0.06\, & 0.27\,$\pm$\,0.03\, & 0.007\,$\pm$\,0.002\, & 1.4   & 1.5   & 2.22\,$^{+1.28}_{-0.77}$ & 0.002    \\
B2   & 1.36\,$\pm$\,0.06\, & 1.34\,$\pm$\,0.06\, & 0.30\,$\pm$\,0.03\, & 0.006*               & 1.4   & 1.4   & 2.09\,$^{+1.00}_{-0.53}$ & 0.03   \\
B3   & 1.36\,$\pm$\,0.06\, & 1.34\,$\pm$\,0.06\, & 0.30\,$\pm$\,0.03\, & 0.006\,$\pm$\,0.002\, & 1.4   & 1.5   & 2.09\,$^{+1.33}_{-0.52}$ & 0.006    \\ \hline
C1   & 1.35\,$\pm$\,0.06\, & 1.33\,$\pm$\,0.06\, & 0.27\,$\pm$\,0.03\, & 0.004\,$\pm$\,0.002\, & 1.4   & 1.5   & 2.32\,$^{+0.32}_{-0.54}$ & 0.05   \\
C2   & 1.31\,$\pm$\,0.06\, & 1.29\,$\pm$\,0.06\, & 0.30\,$\pm$\,0.03\, & 0.006*       & 1.4   & 1.6   & 2.35\,$^{+0.20}_{-0.28}$ & 0.3     \\
C3   & 1.31\,$\pm$\,0.06\, & 1.29\,$\pm$\,0.06\, & 0.30\,$\pm$\,0.03\, & 0.006\,$\pm$\,0.002\, & 1.6   & 1.4   & 2.25\,$^{+0.48}_{-0.27}$ & 0.07    \\  \hline
D1   & 1.31\,$\pm$\,0.06\, & 1.29\,$\pm$\,0.06\, & 0.27\,$\pm$\,0.03\, & 0.005\,$\pm$\,0.002\, & 1.5   & 1.4   & 2.58\,$^{+0.39}_{-0.42}$ & 0.4    \\
D2   & 1.32\,$\pm$\,0.06\, & 1.30\,$\pm$\,0.06\, & 0.30\,$\pm$\,0.03\, & 0.006*               & 1.6   & 1.4   & 2.44\,$^{+0.25}_{-0.20}$ & 0.9   \\
D3   & 1.32\,$\pm$\,0.06\, & 1.30\,$\pm$\,0.06\, & 0.29\,$\pm$\,0.03\, & 0.005\,$\pm$\,0.002\, & 1.6   & 1.4   & 2.52\,$^{+0.31}_{-0.36}$ & 0.7   \\  \hline   
\end{tabular}

\label{tab:results}

\tablefoot{The first column provides the naming of the different optimization, while the second column indicates those parameters included in the \chisq optimization, and the third column indicates any constraints in the model input. The fourth column indicates the common age (Age$_{1}$\,=\,Age$_{2}$) the \chisq minimization provides, including the respective error bars, obtained as described in \ref{sec:error}. The fifth column provides the mass of the primary $M_1$ for the best-fitting model, while the sixth and seventh columns provide the initial metallicity $Z$ and initial helium $Y$ for this model, respectively. The last column indicates the minimum \chisq value obtained for this case. No error bars for metallicity are given in the cases marked with * since, in those cases, we forced the metallicity to be at the value of 0.006.}
\end{table*}

\begin{figure*}
    \centering
    \includegraphics[width=0.8\textwidth,height=70mm
    ]{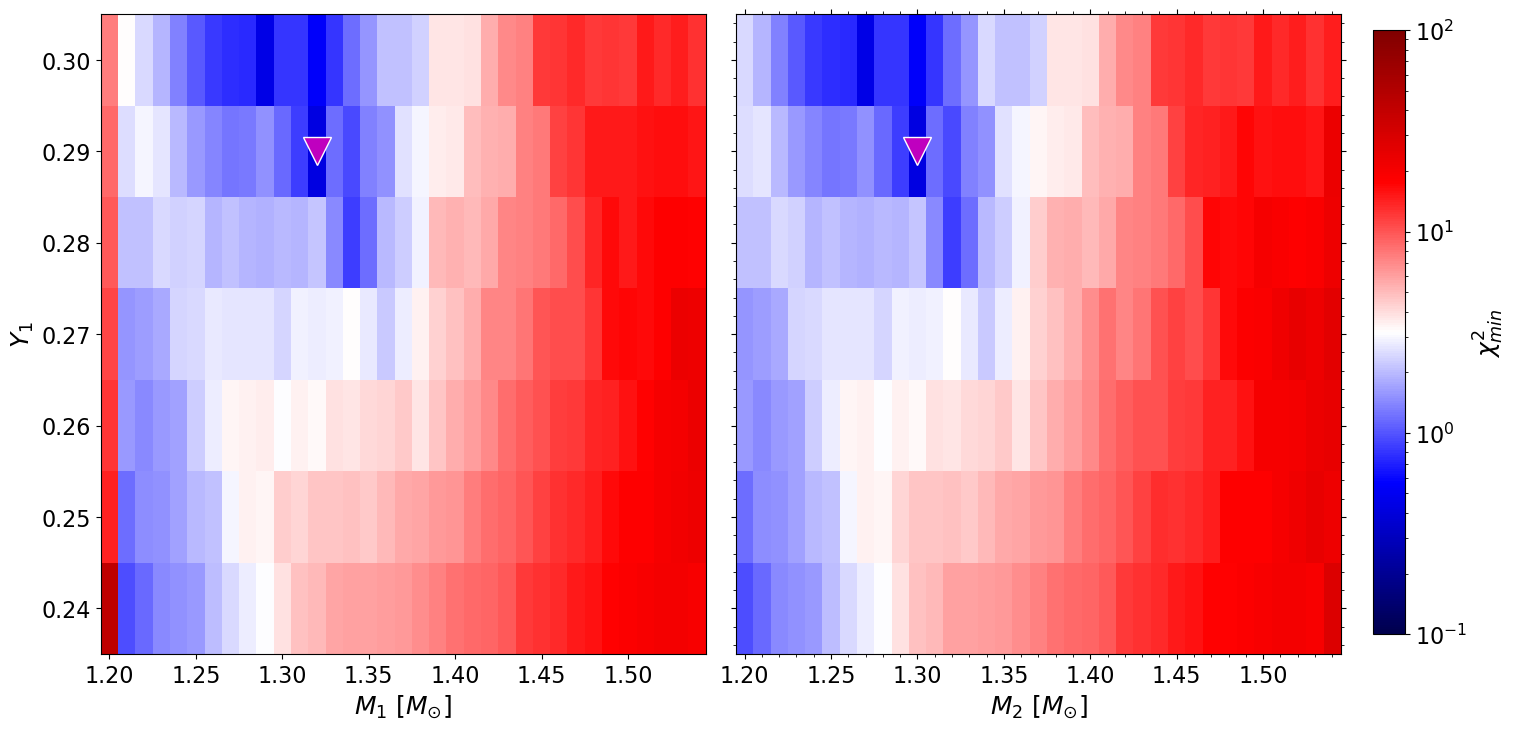}
    \caption{Residuals of \chisq-Minimization in the mass-initial helium plane for both components (primary left, secondary right), for case D3. The best-fitting solution for each is marked as a magenta triangle.}
    \label{fig:massheliumopen}
\vspace{3mm}
    \centering
    \includegraphics[width=0.8\textwidth,height=70mm
    ]{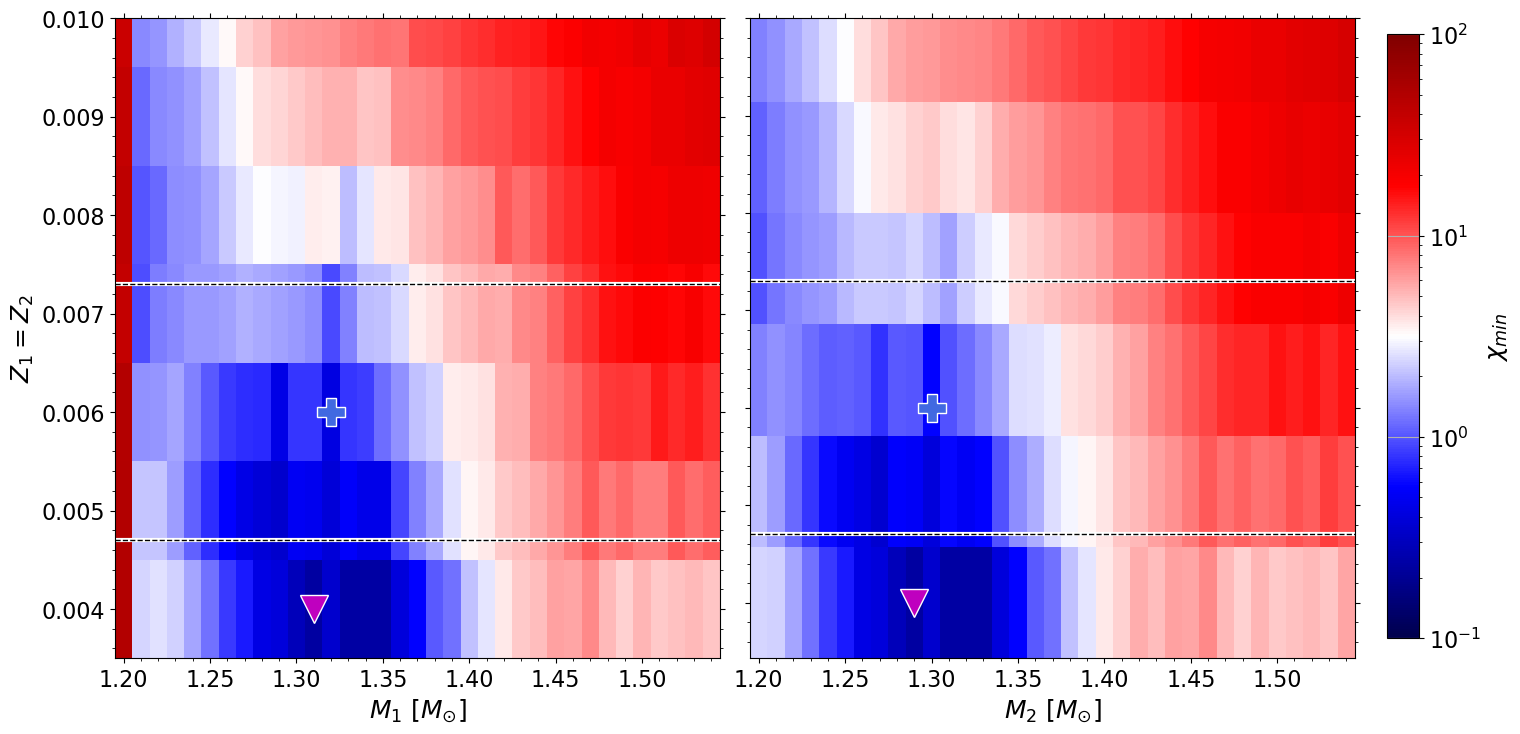}
    \caption{Residuals of the \chisq-minimization in the mass-metallicity plane for both components (primary left, secondary right), for identical metallicities for both components (Case D1). For this case, the magenta triangle marks the best-fitting solution. For the analogous case, with both model metalicities forced to be in the 1-$\sigma$-uncertainty of the spectroscopic value (Case D3), the best fitting solution is given as a blue cross. The dashed lines show the $1\sigma$-errorbar around the spectroscopic metallicity.}
    \label{fig:metallicity}
\end{figure*}

\subsection{Case D: Modeling including magnitude difference}
\label{sec:magdif}
Up to this point, we have used more constraints regarding the primary than the secondary. One additional constraint we could use is the frequency of maximum oscillation \num of the secondary. As described in Sect.\,\ref{sec:observations}, its \num is reflected from the super-Nyquist regime and overlaps with the power excess of the primary and has been estimated using spectroscopic parameters, already included in the figure of merit. Therefore, we refrain from using \num of the secondary as an additional observable.
 The frequency of maximum oscillations \num is correlated with the total luminosity $L$ of a star \citep{Brown1991} by,
 \begin{equation}
\nu_{\text{max}} \propto \frac{M}{L} T_{\text{eff}}^{3.5}.
 \end{equation}
 Because both components are at the same distance from us and affected by the same extinction, we can use the difference between the two absolute magnitudes as a proxy for the luminosity difference, and hence derive \num of the secondary.  The magnitude difference between the two components was inferred by \citetalias{Beck2018} using the spectroscopic parameters (Table \ref{tab:observables}) and the total flux of the system in Johnson V. This led to an estimation of $\Delta$\,$m_v $\,=\,0.58\,$\pm$\,0.08. The theoretical value is calculated as the difference of the default MESA output of magnitudes in Johnson V between the model of the primary and secondary at every timestep.

We, therefore, extend case C with the magnitude difference as an additional parameter. We define case D as,
\begin{equation}
\label{eq:chitot1}
     \chi_{D}^{2} =  \chi_{T_{\text{eff,1}}}^{2} + \chi_{T_{\text{eff,2}}}^{2}+  \chi_{\log g_2}^{2} + \chi_{q}^{2} +  \chi_{\nu_{max,1}}^{2} + 
     \chi_{\Delta\nu_1}^{2} +
     \chi_{\Delta\mathrm{m_V}}^{2}.
\end{equation}

Adding the magnitude difference as an observable results in a distinct distribution of \chisq-minima over age (Fig.\,\ref{fig:agenumax2}). For all cases, the \chisq values are close to 1. The ages of the best fitting model combinations for case D range from 2.44\,Gyr to 2.58\,Gyr, with uncertainties in the same order of magnitude as in case C (Table\,\ref{tab:results}). The masses given by this model case resemble those of cases C2 and C3, with cases D1 being 0.01\,\Msun lower than the masses for cases D2 and D3 ($M_1$\,=\,1.31\,$\pm$\,0.06\,\Msun and $M_2$\,=\,1.30\,$\pm$\,0.06\,\Msun, Fig.\,\ref{fig:dnumassD}).

As another aspect, we consider the variations in initial helium abundance and its effect on the best-fitting model in Fig.\,\ref{fig:massheliumopen}.  For case D3 the models suggest helium abundances $Y_{1,2}$\,=\,0.29. This is also true if we do not force any binary condition regarding the helium abundances.
Regarding the stellar metallicity, Fig.\,\ref{fig:metallicity} depicts an example of the figure of merit as a function of metallicity and mass for the metallicity restricted to case D3. We observe a clear tendency to low metallicity, and the best fitting models all relax around $Z_{1,2}$\,=\,0.005.

\begin{figure*}[t!]
    \centering
    \includegraphics[width=0.85\textwidth]{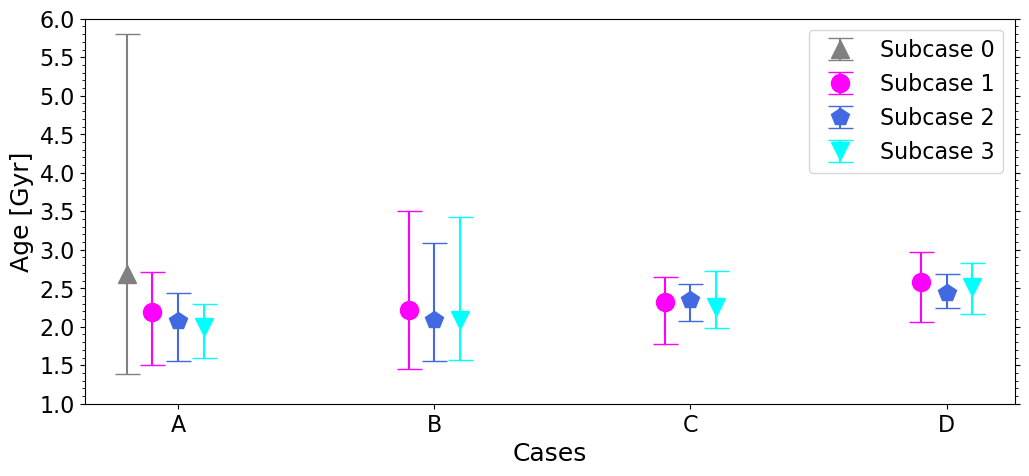}
    \caption{Age estimations for the different cases of the figure of merit, including their uncertainties. The cases are calculated as described in Sect.\,\ref{sec:bestfit}, with the cases color coded as follows:
    magenta: $Z_1$\,=\,$Z_2$; royal blue: $Z_{1,2}$\,=\,$Z_{\mathrm{spec}}$; cyan: $Z_{1,2}$\,=\,$Z_{\mathrm{spec}}$\,$\pm$\,$\sigma_{Z_{\mathrm{spec}}}$. }
    \label{fig:agecases}
\vspace{3mm}
    \centering
    \includegraphics[width=0.85\textwidth]{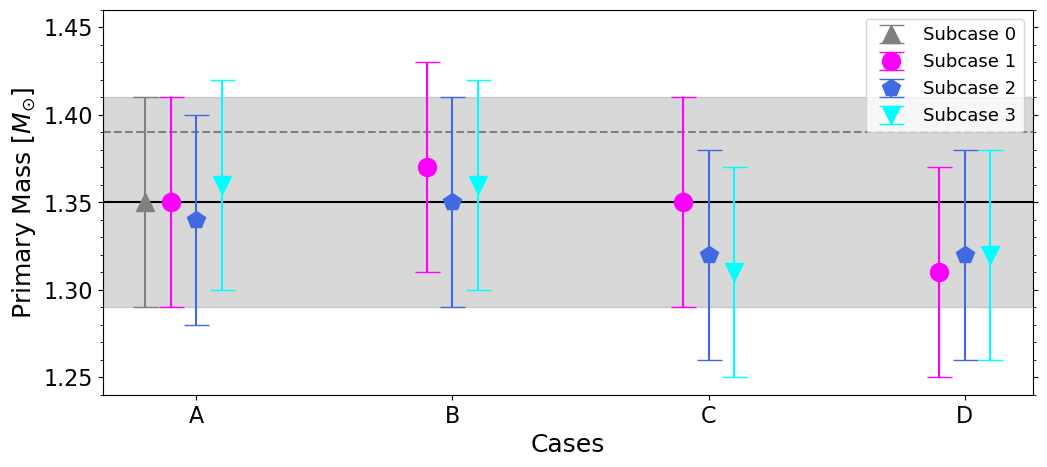}
    \caption{Mass estimations for the primary for different cases of the figure of merit, including their uncertainties. The color coding is the same as in Fig.\,\ref{fig:agecases}. The black solid line indicates the revised mass estimation from observations, with the gray area indicating the respective 1-$\sigma$ error. The gray dashed line indicates the mass as given in \citetalias{Beck2018}}.
    \label{fig:masscases}
\end{figure*}

\section{Discussion} \label{sec:discussion}
Here, we compare the results for the stellar age and fundamental parameters of \KIC derived in the previous section and place them in the context of existing modeling/observational studies in the literature. We focus on cases C and D because these are the most physically motivated ones (Sect.\,\ref{sec:cases}).

\subsection{Stellar age} 
\label{sec:discage}
A comparison of the age determination results for models using the different observables and constraints (cases A to D, including the subcases) is shown in Fig.\,\ref{fig:agecases}. The least constrained case A0 provides an age estimation of 2.68\,Gyr with an absolute age range within the error bars of  $\sigma_t$\,(A0)\,=\,4.4\,Gyr ($\sim$\,100\,\%). The results for this case are similar to a single star or components of an unconstrained binary and serve as a reference point if no additional constraints are available. Applying the binary condition of $Z_1$\,=\,$Z_2$ in cases A1, A2, and A3 reduces the uncertainty in the age of the models and raises the respective \chisq-minima towards one, demonstrating that constraints from binarity help breaking degeneracies in parameter space. For cases B, where \num is applied in addition to the spectroscopic constraints, consistent ages at around 2.0\,Gyr arise from the modeling.  All three subcases of case B exhibit error bars of similar size, indicating that the impact of \num in the figure of merit overwrites the effect of the constraints from binarity. Additionally, we note that the difference in uncertainty between cases A and B could be due to competing minima from pure spectroscopic (cases A) and spectroscopic and asteroseismic (cases B) parameters.

If we move to cases where we use all seismic with all binary constraints the centroids for the ages are slightly higher compared to the previous cases (between 2.25\,Gyr for case C3 to 2.35\,Gyr for case C2), the respective ranges in uncertainties are considerably smaller, with the value for C2, being $\sigma_t$\,(C2)\,=\,0.48\,Gyr ($\sim$10$\%$). Furthermore, similarly to case B, we note that more constraints lead to a \chisq value closer to 1, hence more realistic physical models.

We see a very similar picture for cases D, where the magnitude difference is also included in addition to the spectroscopic and seismic constraints. Case D1 provides the highest age estimation of all binary constrained models with an age of 2.58\,Gyr, with an errorbar range of $\sigma_t$\,(D1)\,=\,0.81\,Gyr. This result can likely be attributed to the \chisq distribution being flat at the minimum point of case D1, with several models with almost identical \chisq values but ages varying from 2.2\,Gyr and 2.6\,Gyr. Case D2 behaves almost identically to case C2 regarding the value and the error bars, suggesting that the metallicity restriction overrides any effects introduced by the magnitude difference. Both subcases (C2 and D2) provide similarly sized uncertainties ($\sigma_t$\,(C2)\,=\,0.48\,Gyr and  $\sigma_t$\,(D2)\,=\,0.45\,Gyr), hence being the only cases with a relative age uncertainty below 10\,\%.  Conclusively, case D3 provides an age of 2.52\,Gyr, with an uncertainty of $\sigma_t$\,(D3)\,=\,0.67\,Gyr, hence about 13\,\% but in the same order of magnitude as for Cases C2 and D2. Cases D2 and D3 are also the most physically motivated ones, as they include all relevant parameters from spectroscopy, asteroseismology, and photometry in addition to reasonable binary constraints regarding the chemical abundances. These cases result in age uncertainties of 9\,\% and 13\,\%, respectively.

The best fitting ages determined from all cases, including seismic constraints (B, C, D), lead to consistent age determinations with scatter of about 12\,\%. As visible from Fig.\,\ref{fig:agecases}, the cases including \dnu as a constraint (case C and D) provide ages that are $\sim$\,0.4\,Gyr higher compared to respective subcases of cases A and B. Conclusively, all ages determined agree within uncertainties.

\subsection{Stellar mass}
The masses determined from the modeling for the different observables and constraints are depicted in Fig.\,\ref{fig:masscases}. It can be seen that the masses for cases C and D (including the constraint of \dnu) are on average around 5\,\% lower than those for cases A and B (no \dnu constraint).
To consistently compare the masses from modeling with the observed seismic masses, we need to redetermine the masses from the scaling relations, taking the correction factor $f_{\Delta\nu}$ introduced in Sect.\,\ref{sec:seism} into account. Using Eq.\,14 from \cite{Li2023Surface} and a $f_{\Delta\nu}$\,=\,0.974, derived from the centroids of the observational parameters of the primary and Eq.\,16 from \cite{Li2023Surface}, leads to a revised primary mass of $M_1$\,=\,1.35$\,\pm$\,0.06\,\Msun and, using the mass ratio from spectroscopy, a secondary mass of $M_2$\,=\,1.33\,$\pm$\,0.06\,\Msun.

All values from our modeling fall well within the 1-$\sigma$ uncertainties of the scaling-based mass of both components (Fig.\;\ref{fig:masscases}) and also agree with one another within uncertainties. This is an important indication of our model correctly reproducing the two stellar components of \KIC.
When including the local \dnu in the figure of merit in case C, we observe slightly lower masses for the more constrained cases C2 and C3, with primary masses for those cases being 1.31\,\Msun. For case D, where we include the magnitude difference into the figure of merit, we receive similarly low masses as in case C, with the primary mass between 1.31\,\Msun to 1.32\,\Msun and the secondary mass between 1.29\,\Msun to 1.30\,\Msun. This behavior indicates that mass estimations, including \num and \dnu, tend to provide slightly lower mass models than results solely from spectroscopy. Eventually, this is consistent with the higher ages obtained for cases C and D discussed in the previous subsection (Fig.\,\ref{fig:agecases}). The mass consistency among cases C and D aligns with those cases being the most physically motivated ones.

As shown in Fig.\,\ref{fig:masscomparison}, \ref{fig:dnumass} and \ref{fig:dnumassD} the mass ratio is clearly reproduced when the residuals are plotted in the $M_1$-$M_2$ plane. Experiments have shown that this is also the case if $q$ is not included in the figure of merit. For all cases, invariant of constraints or different figures of merit, the mass ratio $q$\,=\,1.015 is always conserved, hence pointing to the mass ratio as a unique benefit and reliable constraint arising from the SB2 nature of the system.

\subsection{Convective mixing length}
 The exact value for $\alpha_{\mathrm{MLT}}$ varies between the cases without any obvious correlations (Table\,\ref{tab:parameters}). This behavior might be caused by the model correcting itself for possible systematic problems in either of the surface temperature measurements. However, the $\alpha_{\mathrm{MLT}}$ for the primary is lower or equal to the $\alpha_{\mathrm{MLT}}$ of the secondary for most of the cases, except case D. This pattern of $\alpha_{\mathrm{MLT,1}}$\,$\leq$\,$\alpha_{\mathrm{MLT,2}}$ indicates that the $\alpha_{\mathrm{MLT}}$ might not be constant over the evolution of a star, but rather it could decrease as the star rises on the sub-giant/red-giant branch
when the extension on the super-adiabatic region strongly increases.  This result appears to provide some support to previous mixing-length calibration studies, such as \cite{JoyceTayar2023,Trampedach2014ML}, as well as it shows that relying on a single solar-calibrated $\alpha_{\mathrm{MLT}}$ value in stellar modeling might not be fully advisable, especially for stars whose
mass, evolutionary stage and/or chemical composition are significantly different than those of Sun-like stars.  For instance, \cite{Tayar2017MLTMetal} suggested that the super-adiabatic convection efficiency
should decrease with decreasing metal abundance, but on the same topic \cite{Salaris2018} - by using a different set of model predictions - have obtained a quite different conclusion, and have shown
that the exact heavy element distribution (solar scaled versus $\alpha-$enhanced one) plays a role in the comparison between observations and stellar models. However, as concluded in \cite{JoyceTayar2023} and from results of synthetic studies \citep[e.g. ][]{Valle2019}, any direct dependence of mixing length on metallicity should be dealt with caution. Needless to say, to
shed light on this important topic, it is important to increase the sample of (binary) stars with accurately determined physical properties.

As we have limited the range for $\alpha_{\mathrm{MLT}}$ as described in Appendix\,\ref{sec:alphamlt} using the spectroscopic case A, we retested if, for the Cases B to D, better fitting models occur outside the limited range. This test confirmed our choice of limiting $\alpha_{\mathrm{MLT}}$, as no best-fitting models outside the range were found.

\subsection{Initial He and metallicity}
\label{sec:helium}
Regarding the initial helium abundance, there is a distinct trend. Without exception, all cases show an initial helium abundance greater than the generally accepted primordial helium abundance of 0.24 \citep{Charbonnel2017}. The modeled helium abundances range from 0.27 to 0.30, with a formal uncertainty of $\pm$\,0.03. Due to the comparatively young age of the system of around 2\,Gyr, born around 11\,Gyr after the Big Bang, this enriched initial helium value is quite plausible and also consistent with the modeling of similar - although single - stars \citep{Verma2019}. Furthermore, enforcing stronger metallicity constraints, as in subcases 2 and 3, respectively, leads to higher initial helium abundances than those with less stringent or no constraints (subcases 0 and 1). As some of the recovered initial helium values are located at the edge of the explored range, we explored the modeling grid with a higher upper boundary of $Y_{\mathrm{ini}}$\,=\,0.33. The majority of results results were unchanged with this expanded range. However, for two subcases (C2 and D2) we recovered higher abundances, again at the edge of the explored range ($Y_{\mathrm{ini}}$\,=\,0.32 and $Y_{\mathrm{ini}}$\,=\,0.33, respectively). This edge effect could be related to the effect recovered from synthetic modeling by \cite{Valle2024}, who demonstrate that, independent of the fitting method, helium abundances in grid modeling are biased towards the edges of the explored ranges.

Comparing the results for the metallicity of the best-fitting models, it occurs that subcases 1, with no numerical restriction for the initial heavy element abundance, exhibit a $Z$ value at the lower boundary of the model grid range ($Z_{1,2}$\,=\,0.004\,$\pm$\,0.002). This is significantly lower than the spectroscopic value for $Z$, even when accounting for the observational errors (Fig.\,\ref{fig:metallicity}). Also, case A0, with no binarity constraints, exhibits an initial heavy metal abundance of $Z_{1,2}$\,=\,0.004\,$\pm$\,0.002. Following directly from the constraints in cases 2 and 3, forcing both components to the spectroscopic and spectroscopic metallicity, including the 1-$\sigma$ observational error bars, respectively, lead to higher metallicities, as shown in Table\,\ref{tab:results}. Apart from this trend, we observe no clear correlations in the results for metallicity. However, it is a crucial parameter for constraining the model parameter space, particularly for the age determination, as we have demonstrated in Sect.\,\ref{sec:discage}.
\\

\subsection{Figures of merit and model uncertainty}

Eventually, it should be noted that - following the description in Sect.\,\ref{sec:error} - all uncertainties for stellar parameters, except the age, are dominated by the observational uncertainties of the input parameters. Therefore, an improvement in the uncertainty of those parameters would directly lead to an improvement in the model uncertainties. Across all cases, subcase 2 systematically has smaller uncertainties because it is chosen for a single metallicity bin, while subcases 3 take a wider bin and therefore allows for more combinations, which increases the uncertainty.
As cases D demonstrate the most consistent age estimations, and exhibit a \chisq closest to unity, and include all relevant parameters from spectroscopy, asteroseismology, and photometry in addition to reasonable binary
constraints, we consider them to be the most physically motivated ones.

\section{Conclusion} \label{sec:conclusion}

In this work, we performed comprehensive modeling of the red giant-subgiant asteroseismic SB2 binary system \KIC. We build a multidimensional grid in parameter space to accurately model the primary and secondary using the MESA stellar evolution code and GYRE oscillations code. By using constraints from spectroscopy, asteroseismology, and binarity and different formulations of a \chisq figure-of-merit, we determined precise ages and stellar parameters ($M$, $Z$, $Y$, $\alpha_{\mathrm{MLT}}$) for the binary system. These results are summarized in Table\,\ref{tab:results}. From the results of the most physically motivated cases (Case D2 and D3, see Sect.\,\ref{sec:discage}) we adopt 2.44\,$^{+0.25}_{-0.20}$\,Gyr (Case D2) as the age of the system.

All cases with binarity constraints tested in this work converge to ages within less than 20 \% of each other, with a maximum of 2.58\,Gyr (Case D1) and a minimum of 2.09\,Gyr (Case B2 and B3), while those cases including all seismic constraints (C, D) lead to consistent age determinations as well (scatter of less than 13\,\%). This demonstrates the robustness of age determination using the combined approach with data from spectroscopy, asteroseismology, and binarity. Additionally, we showed that, in particular, constraints from binarity lead to a significant improvement in age uncertainties, for the best case bringing them down to 9\,\%, an order of magnitude improvement compared to a modeling approach, without asteroseismic and binarity constraints (50-100\,\%, see Table \ref{tab:results}). Furthermore, across all modeling cases, the minimum \chisq values get closer to 1 when applying constraints from binarity, pointing to less overfitting and more realistic models. In addition, our analysis also demonstrates that adequately calibrated parameters from asteroseismology, in particular \dnu, can lead to more precise age estimations compared to using only observables from spectroscopy (Fig.\,\ref{fig:agecases}). We conclude that constraints from binarity and asteroseismology can break degeneracies and certain limitations arising from 1D-modeling \citep[e.g., see conclusions of ][]{JoyceTayar2023} and improve the precision of age determination significantly.

With this detailed study, we demonstrated that well-modeled systems like \KIC, which are observationally well constrained from photometry, spectroscopy, and asteroseismology, have the diagnostic potential for testing the internal physics that otherwise is challenging from single star models. We showed this by constraining the initial helium abundance $Y$ and the mixing-length parameter $\alpha_{\mathrm{MLT}}$, both parameters usually inaccessible to observations. The higher than primordial initial helium abundance and lower than solar $\alpha_{\mathrm{MLT}}$ shown in our modeling is compatible with the system's age and evolutionary stage and in agreement with previous studies. However, as discussed in Sect.\,\ref{sec:helium}, some of the recovered initial helium values could be affected by a general bias towards the grid edges \citep{Valle2024}.

 This in-depth modeling demonstrated that \KIC is a benchmark system for constraining stellar age. As stated in \cite{Miglio2014}, many constrained systems like the prototype \KIC should exist; however, few of them have been found, none of which has been modeled in such depth. Further analysis is necessary to find more oscillating binary systems like \KIC with similar constraints from seismology and binarity. 

Possible further constraints, particularly for age determination, could arise from the study of eclipsing binary systems \citep{Gaulme2022,Valle2023a,Rowan2024} or stellar clusters \citep{Brogaard2023, Reyes2024} containing solar-like oscillators.
\cite{BeckGrossmann2023} has reported over 900 new oscillating binary systems by cross-correlation of \Gaia DR3 data with sample catalogs of solar-like oscillators. The authors suggest that astrometric binary systems for which an SB2 signature has been found could significantly expand the sample size of benchmark systems like \KIC in number, evolutionary states, and parameter space. Also, using individual frequencies of the modes in future modeling can help to improve further the estimation of stellar ages and other parameters \citep{LiTanda2022,LiTanda2024}.

Future asteroseismic space missions, particularly the ESA PLATO Mission \citep{Rauer2014,Rauer2024}, will be able to provide us with high signal-to-noise asteroseismic data of thousands of unstudied main-sequence and red-giant oscillators. In particular, the Science Validation and Calibration PLATO Target Input Catalogue (scvPIC) contains several thousand potential benchmark binary systems with known inclinations, which potentially host main-sequence or giant solar-like oscillators. Using these future datasets in combination with existing ones, such as The Apache Point Observatory Galactic Evolution Experiment \citep[APOGEE; ][]{APOGEE} and Gaia \citep{GaiaMain}, and extended by ground-based follow-up programs will enable us to use constraints from asteroseismology, spectroscopy and binarity to advance our knowledge about the Milky way's history, dynamics and chemical composition in the context of galactic archaeology.
\vspace{5mm}
\begin{acknowledgements}

We thank the referee, Pier Giorgio Prada Moroni, for constructive comments that improved the paper.
The authors thank the people behind the ESA \Gaia, NASA \Kepler, and NASA TESS missions and HERMES spectrograph.

The authors thank The team of UniIT, especially Dr. Ursula Winkler and David Bodruzic is thanked for excellent support and maintanance of the High Performance Computing cluster of the University of Graz (Graz Scientific Cluster 1 - GSC 1), used for computations of this work. 
Further, thanks go to Roland Maderbacher and Klaus Huber for their support in high-performance computing. We also thank Anna Querioz and Carlos Allende for the discussions that helped improve the target parameters.\\

The project that gave rise to these results received the support of a fellowship from ”la Caixa” Foundation (ID 100010434). The fellowship code is LCF/BQ/DI23/11990068.
DHG acknowledges support of the Dr. Heinrich-Jörg Foundation at the Graz University. 
DHG and NM acknowledge travel support through the "Förderungsstipendium" provided by the Faculty of Natural Sciences of the Graz University.
PGB acknowledges support by the Spanish Ministry of Science and Innovation with the \textit{Ram{\'o}n\,y\,Cajal} fellowship number RYC-2021-033137-I and the number MRR4032204. 
PGB, DHG, DGR, and RAG acknowledge support from the Spanish Ministry of Science and Innovation with the grant no. PID2023-146453NB-100 (\textit{PLAtoSOnG}). DHG, SM, DGR, and RAG acknowledge support from the Spanish Ministry of Science and Innovation with the grant no. PID2023-149439NB-C41 (\textit{PLATO}).
SM acknowledges support by the Spanish Ministry of Science and Innovation with the grant no. PID2019-107061GB-C66 and through AEI under the Severo Ochoa Centres of Excellence Programme 2020--2023 (CEX2019-000920-S).
SM and DGR 
acknowledges support from the Spanish Ministry of Science and Innovation with the grant no. PID2019-107187GB-I00.
RAG acknowledges support from the PLATO Centre National D'{\'{E}}tudes Spatiales grant. DGR acknowledges support from the Juan de la Cierva program under contract JDC2022-049054-I. 
LS acknowledges the Graz University of Technology travel grant.
SC has been funded by the European Union – "NextGenerationEU" RRF M4C2 1.1 n: 2022HY2NSX. "CHRONOS: adjusting the clock(s) to unveil the CHRONO-chemo-dynamical Structure of the Galaxy” (PI: S. Cassisi), and INAF 2023 Theory grant "Lasting" (PI: S. Cassisi). We gratefully acknowledge support from the Australian Research Council through Laureate Fellowship FL220100117, which includes a PhD scholarship for LSS.

This work has made use of data from the European Space Agency (ESA) mission
\Gaia (\url{https://www.cosmos.esa.int/gaia}), processed by the \Gaia
Data Processing and Analysis Consortium (DPAC,
\url{https://www.cosmos.esa.int/web/gaia/dpac/consortium}). Funding for the DPAC
has been provided by national institutions, particularly the institutions
participating in the \Gaia Multilateral Agreement. 
This paper includes data collected with the \textit{Kepler}\,\&\,\TESS missions obtained from the MAST data archive at the Space Telescope Science Institute (STScI). 
Funding for these missions is provided by the NASA Science Mission Directorate and the NASA Explorer Program. STScI is operated by the Association of Universities for Research in Astronomy, Inc., under NASA contract NAS 5–26555.
This paper is partly based on observations obtained with the HERMES spectrograph, which is supported by the Research Foundation - Flanders (FWO), Belgium, the Research Council of KU Leuven, Belgium, the Fonds National de la Recherche Scientifique (F.R.S.-FNRS), Belgium, the Royal Observatory of Belgium, the Observatoire de Genève, Switzerland and the Thüringer Landessternwarte Tautenburg, Germany.\\
\textit{Software:} \texttt{Python} \citep{10.5555/1593511}, 
\texttt{numpy} \citep{numpy,Harris_2020},  
\texttt{matplotlib} \citep{4160265},  
\texttt{scipy} \citep{2020SciPy-NMeth},
\texttt{Astroquery} \citep{Ginsburg2019}.
This research made use of \texttt{Astropy} \citep{astropy:2013, astropy:2018}, a community-developed core Python package for Astronomy. This work has utilized the stellar evolutionary code package, Modules for Experiments in Stellar Astrophysics
\citep[MESA][]{Paxton2011, Paxton2013, Paxton2015, Paxton2018, Paxton2019, Jermyn2023}. The MESA EOS is a blend of the OPAL \citep{Rogers2002}, SCVH
\citep{Saumon1995}, FreeEOS \citep{Irwin2004}, HELM \citep{Timmes2000},
PC \citep{Potekhin2010}, and Skye \citep{Jermyn2021} EOSes.
Radiative opacities are primarily from OPAL \citep{Iglesias1993,
Iglesias1996}, with low-temperature data from \citet{Ferguson2005}
and the high-temperature, Compton-scattering dominated regime by
\citet{Poutanen2017}.  Electron conduction opacities are from
\citet{Cassisi2007} and \citet{Blouin2020}.
Nuclear reaction rates are from JINA REACLIB \citep{Cyburt2010}, NACRE \citep{Angulo1999} and
additional tabulated weak reaction rates \citet{Fuller1985, Oda1994,
Langanke2000}.  Screening is included via the prescription of \citet{Chugunov2007}.
Thermal neutrino loss rates are from \citet{Itoh1996}. This paper utilized the GYRE stellar oscillations code developed by \cite{Townsend2013}.
\end{acknowledgements}

%
\bibliographystyle{aa} 
\bibliography{bib-refs.bib} 
%
\begin{appendix}

\section{Adjustment of $\alpha_\mathrm{MLT}$}
\label{sec:alphamlt}

 As recently summarized by \cite{JoyceTayar2023}, the application of one-dimensional mixing-length theory in 1D modeling and, in particular, the choice of $\alpha_{\mathrm{MLT}}$ is a complex topic and can have a significant impact on modeled fundamental parameters and degeneracies with other physical parameters and also the stellar age. For computational efficiency, we decided to narrow down the range of $\alpha_{\mathrm{MLT}}$ before proceeding with the investigation. We have decided to calibrate the mixing length by testing the impact of a changing $\alpha_{\mathrm{MLT}}$ on the best-fitting models of the previously introduced spectroscopic case. 
 
\begin{figure}
    \centering
    \includegraphics[width=\columnwidth
     ]{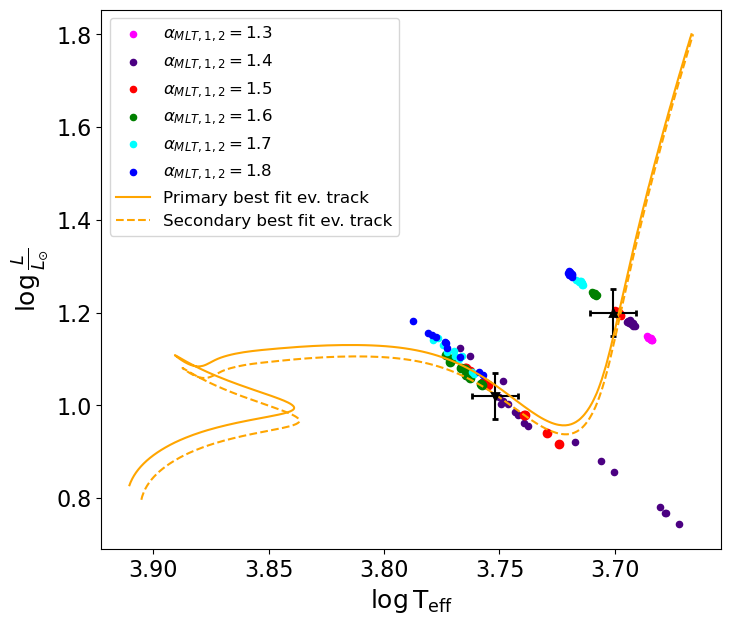}
    \caption{Position of the 10 best models for the spectroscopic case for different choices of alpha concerning the observations, with the observational values including error bars given as black triangles. The evolutionary tracks of the primary and secondary for the best-fitting case are given in (dashed) orange lines. These two evolutionary tracks correspond to models with $\alpha_{\mathrm{MLT}}$\,=\,1.5, $Z$\,=\,0.004 and masses of $M_1$\,=\,1.39\,\Msun and $M$\,=\,1.37\,\Msun.}
    \label{fig:hrdalpha}
\end{figure}

For the adjustment of the $\alpha_{\mathrm{MLT}}$ we have used the figure of merit with parameter set from case A1 (Sect.\,\ref{sec:spec}) as a benchmark. 
  Fig.\,\ref{fig:hrdalpha} depicts the position of the 10 model combinations with the lowest \chisq for each choice of $\alpha_{\mathrm{MLT}}$ on a HRD-like plane. We observed that even the best fitting models for the spectroscopic case with the solar calibrated value $\alpha_{\mathrm{MLT}}$\,=\,1.8 for the primary results in significantly higher luminosity and temperature compared to the spectroscopic values. In consistency with literature, a lower $\alpha_{\mathrm{MLT}}$ decreases both the temperature and luminosity \citep{JoyceTayar2023}, causing a significant shift of position in the HRD. For $\alpha_{\mathrm{MLT}} = 1.5$ the agreement in HRD-position is best for the primary, with the models for $\alpha_{\mathrm{MLT}} = 1.4$ and $\alpha_{\mathrm{MLT}} = 1.6$ located within the 1-$\sigma$ errorbars of the observations. We observe a similar but more loose trend by varying the $\alpha_{\mathrm{MLT}}$ for the secondary instead. This behavior is expected as the primary is much more sensitive to changes in $\alpha_{\mathrm{MLT}}$ than the secondary (Sect.\,\ref{sec:introduction}). Consistently, our testing showed that the exact choice of $\alpha_{\mathrm{MLT}}$ does not impact the modeling nearly as much as for the primary. Therefore, to reduce dimensionality and avoid further degeneracies, we reduce the $\alpha_{\mathrm{MLT}}$ range for the primary and secondary to values of 1.4, 1.5, and 1.6. Although it is likely that the $\alpha_{\mathrm{MLT}}$ does not change significantly between the two respective HRD-positions \citep{Reyes2024}, we do not restrict the models too rigorously and account for small changes in $\alpha_{\mathrm{MLT}}$ by not forcing the same value for both components.
 
This set of values for $\alpha_{\mathrm{MLT}}$ is lower than the values suggested by, e.g., \cite{Trampedach2014ML} for models with similar mass at a similar evolutionary stage. However, \cite{Trampedach2014ML} assumed solar metallicity for all models, while our target has sub-solar metallicity. This difference in metallicity and the known degeneracy between $\alpha_{\mathrm{MLT}}$ and metallicity \citep{JoyceTayar2023} are a probable cause for the shift in $\alpha_{\mathrm{MLT}}$ between this work and the models from \cite{Trampedach2014ML}.  The analysis of M67 red giants by \cite{Reyes2024} also supports this conclusion.

\section{Details on modeling}

\subsection{GYRE example inlist} \label{app:GYRE}
In Listing\,\ref{list:GYRE} we provide an example inlist for the computation of the individual frequencies of the models using GYRE.
\begin{lstlisting}[language=Fortran,
    caption = GYRE example inlist, 
    label= list:GYRE,frame=Tb]
&constants
/

&model
  model_type = 'EVOL'
  file = 'profile100.data.FGONG'
  file_format = 'FGONG'
/

&mode
  l = 0
/


&osc
  outer_bound = 'ISOTHERMAL'
/

&rot
/

&num
  diff_scheme = 'COLLOC_GL4'
/

&scan
  grid_type = 'LINEAR'
  freq_min = 1
  freq_max = 3000
  n_freq = 5000
  freq_min_units = 'UHZ'
  freq_max_units = 'UHZ'
/

&grid
  w_osc = 2
  w_exp = 1
  w_ctr = 4
/


&ad_output
  summary_file = 'summary100.h5'
  summary_item_list = 'l,n_pg,n_p,n_g, &
  freq, freq_units,E_norm, M_star, &
  R_star, L_star'
  freq_units = 'UHZ'
/

&nad_output
/
\end{lstlisting}

\subsection{MESA example inlist} \label{app:MESA}

In Listing\,\ref{list:MESA}, we show an example MESA inlist for a model with a mass of 1.39\,\Msun, metallicity of $Z$\,=\,0.015 and helium abundance of $Y$\,=\,0.27. The parameters marked in the example as 'variable' were varied from model to model according to the ranges and stepsizes described in Sect.\,\ref{sec:grid} and summarized in Table \ref{tab:parameters}. The remaining parameters were kept constant for all computed models. For concise display, some lines in the inlist were broken with the Fortran line break \&.   

\begin{lstlisting}[language=Fortran,
    caption = MESA example inlist, 
    label= list:MESA,frame=Tb]

&star_job

      history_columns_file = &
      history_columns_list_path
      show_log_description_at_start = .false.
      create_pre_main_sequence_model = .true.

      save_model_when_terminate = .true.
      save_model_filename = 'RGB.mod'
      required_termination_code_string &
      = 'log_L_upper_limit'
      color_num_files = 2
      color_file_names(1) = 'lcb98cor.dat'
      color_num_colors(1) = 11

      color_num_files = 2
      color_file_names(2) = colorfilepath
      color_num_colors(2) = 17

      pgstar_flag = .false.
      pause_before_terminate = .false.

      relax_to_this_tau_factor = 0.001
      relax_tau_factor = .true.
/ ! end of star_job namelist

&eos

/ ! end of eos namelist

&kap

      kap_file_prefix = 'gs98'
      use_Type2_opacities = .true.

/ ! end of kap namelist

&controls

      x_ctrl(1) = 6000d0 ! Teff treshold
      x_ctrl(2) = 1d9 ! Age_treshold
      x_ctrl(3) = 2d5 ! max_timestep_years
      x_ctrl(4) = 1.44 ! mass treshold

      log_L_upper_limit = 1.6d0  !

      mixing_length_alpha = 1.8
      initial_mass = 1.39 ! variable
      initial_z = 0.015 ! variable
      initial_y = 0.27 ! variable
      num_trace_history_values = 2
      trace_history_value_name(1) = &
      'rel_E_err'
      trace_history_value_name(2) = &
      'log_rel_run_E_err'

      cool_wind_full_on_T = 9.99d9
      hot_wind_full_on_T = 1d10
      cool_wind_RGB_scheme = 'Reimers'
      RGB_to_AGB_wind_switch = 1d-4
      Reimers_scaling_factor = 0.2

      varcontrol_target = 1d-3
      delta_lgL_He_limit = 0.01d0
      time_delta_coeff = 0.5

     !mesh
      mesh_delta_coeff = 1
      mesh_min_dr_div_cs = -1
      max_allowed_nz = 30000
      use_other_mesh_delta_coeff_factor = &
      .false.

     !output
      photo_interval = -1
      profile_interval = 4
      history_interval = 1
      terminal_interval = 100
      write_header_frequency = 50

      atm_T_tau_opacity = 'varying'
      atm_T_tau_relation = 'Eddington'

      !overshooting
      overshoot_scheme(1) = 'exponential'
      overshoot_zone_type(1) = 'any'
      overshoot_zone_loc(1) = 'any'
      overshoot_bdy_loc(1) = 'any'
      overshoot_f0(1) = 0.005
      overshoot_f(1) = 0.02
     

      max_num_profile_models = 10000
      max_model_number = 50000

/ ! end of controls namelist



&pgstar


/ ! end of pgstar namelist
\end{lstlisting}

\subsection{Investigation of temporal resolution in MESA \& GYRE} \label{app:resolution} 

\begin{figure}[t!]
    \centering
    \includegraphics[width=\columnwidth,height=100mm]{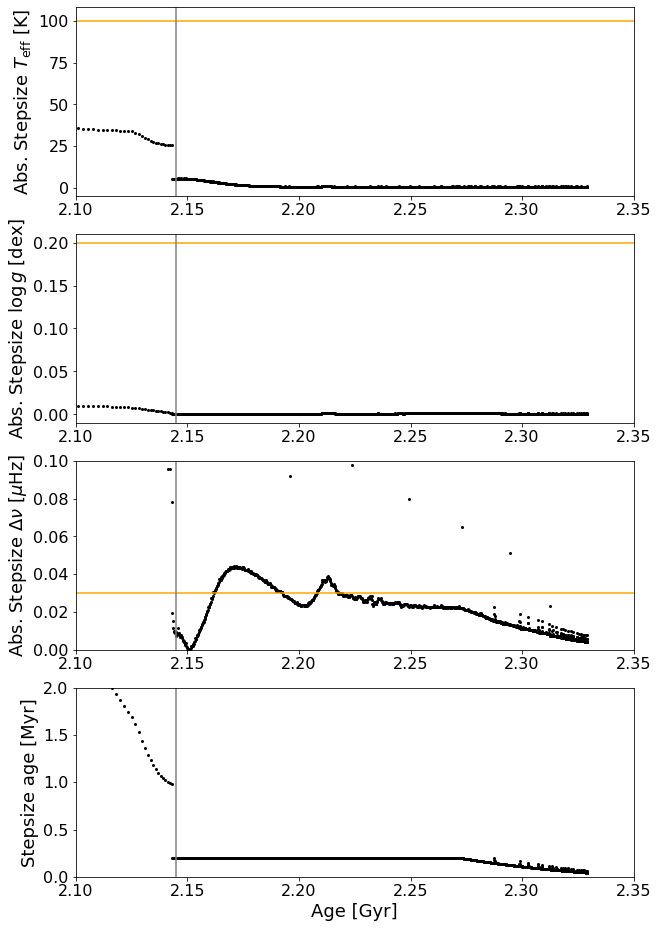}
    \caption{Variation of the parameters used in the figure of merit between consecutive timesteps of a model for a star with 1.39 \Msun and $Z$\,=\,0.007. The top, middle, and bottom panels show the differential evolution of the effective temperature $\mathrm{T_{eff}}$, surface gravity $\log \mathrm{g}$  and \dnu, respectively. The vertical grey line indicates when a maximum timestep of 0.2\,Myr was enforced between two models. The horizontal orange lines indicate the respective observational uncertainties for the given parameters. The bottom panel gives the timestep between consecutive models as a function of age.}
    \label{fig:dnucomp200}
\end{figure}

As described in Sect.\,\ref{sec:seism}, adjusting the timesteps in MESA to provide sufficient input files for calculating \dnu from with GYRE was necessary. The \texttt{time\_delta\_coeff} was set to 0.5, half of its default value for all model runs. This was sufficient to have small enough parameter steps in $\mathrm{T_{eff}}$ and $\log \mathrm{g}$ not to exceed the respective observational uncertainties. However, the steps in \dnu were still orders of magnitudes above the observational uncertainty of 0.03\,$\mu \mathrm{Hz}$. To solve this problem, we needed to find the right trade-off between a small enough timestep to get this desired resolution but not too small to become numerically unstable and require excessive computation time. This trial and error search resulted in a maximum allowed timestep in MESA of 0.2\,Myr enforced from the subgiant branch onward for models cooler than 6000\,K. The differential variations for spectroscopic parameters and \dnu for an exemplary model of 1.39 \Msun and $Z$\,=\,0.007 enforcing the condition mentioned above is shown in Fig.\,\ref{fig:dnucomp200}. While this method works fine for stars up to around 1.44\,\Msun, the differential stepsizes in \dnu increase for masses above it. We therefore implemented another threshold at 1.44\,\Msun, where a maximal step size of 0.1\,Myr was enforced. The necessary modifications in the runstarextra file are shown in Listing\,\ref{list:Runstar}, while the thresholds described above were defined in the respective MESA inlists.

\begin{lstlisting}[language=Fortran,
    caption = Modified extra finish step function in the run\_star\_extra.f90 file., 
    label= list:Runstar,frame=Tb]

integer function extras_finish_step(id)
 use chem_def
 integer, intent(in) :: id
 integer :: ierr
 real(dp) :: Teff_treshold
 real(dp) :: Age_treshold
 real(dp) :: max_timestep_years
 real(dp) :: mass_treshold
 type (star_info), pointer :: s
 ierr = 0
 call star_ptr(id, s, ierr)
 if (ierr /= 0) return
 extras_finish_step = keep_going

 Teff_treshold = s% x_ctrl(1)
 Age_treshold = s% x_ctrl(2)
 max_timestep_years = s% x_ctrl(3)
 mass_treshold = s% x_ctrl(4)


 s% xtra(1) = s% Teff

 if ((s% xtra(1) < Teff_treshold) &
 .AND. (s % star_age > Age_treshold)) then

    s% need_to_update_history_now = .true.
    s% need_to_save_profiles_now = .true.

    s% max_years_for_timestep = &
    max_timestep_years

 endif
 if ((s% initial_mass &+
 .GE. mass_treshold)) then
    s% max_years_for_timestep = &
    max_timestep_years/2



 endif

end function extras_finish_step
\end{lstlisting}

\end{appendix}
\clearpage

\clearpage

\end{document}